\documentclass[twocolumn,amsmath,amssymb,nofootinbib]{revtex4}
\usepackage{graphicx}

\newcommand{\eqeqref}[2]{(\ref{#1}, \ref{#2})}
\newcommand{\eqtoeqref}[2]{(\ref{#1}-\ref{#2})}

\begin{document}
\bibliographystyle{apsrev}
\title{Self-diffusion in granular gases: Green-Kubo versus Chapman-Enskog}
\author{Nikolai V. \surname{Brilliantov}}
\affiliation{Institute of Physics, University Potsdam, Am Neuen Palais 10, 14469 Potsdam, Germany}
\author{Thorsten \surname{P\"oschel}}
\affiliation{Kavli Institute for Theoretical Physics, University of California, Santa Barbara, CA 93106, USA
and\\
Charit\'e, Institut f\"ur Biochemie, Monbijoustra{\ss}e 2, 10117 Berlin, Germany}

\date{\today} 

\begin{abstract}
We study the diffusion of tracers (self-diffusion) in a homogeneously cooling gas of dissipative particles, using the Green-Kubo relation and the Chapman-Enskog approach. The dissipative particle collisions are described by the coefficient of restitution $\varepsilon$ which for realistic material properties depends on the impact velocity. First, we consider self-diffusion using a constant coefficient of restitution, $\varepsilon=$const, as frequently used to simplify the analysis. Second, self-diffusion is studied for a simplified (stepwise) dependence of $\varepsilon$ on the impact velocity. Finally, diffusion is considered for gases of realistic viscoelastic particles. We find that for $\varepsilon=$const  both methods lead to the same result for the self-diffusion coefficient. For the case of impact-velocity dependent coefficients of restitution, the Green-Kubo method is, however, either restrictive  or too complicated for practical application, therefore we compute the diffusion coefficient using the Chapman-Enskog method. We conclude that in application to granular gases, the Chapman-Enskog approach is preferable for deriving kinetic coefficients. 
\end{abstract}

\pacs{PACS}

\maketitle
{\bf The diffusion of tracer particles (self-diffusion) in a force-free gas of dissipatively colliding particles, also called granular gases, is studied. Main theoretical approaches for the calculation of transport coefficients are the Green-Kubo and Chapman-Enskog methods. The former is based on fluctuation-dissipation relations, which express transport coefficients in terms of the time correlation function of dynamical variables. The latter approach is based on the Boltzmann equation, that describes the evolution of the distribution function of tracers. The dissipative nature of the gas is characterized by the coefficient of restitution, $\varepsilon$ which is the central quantity in the theory of granular gases. This coefficient quantifies the loss of kinetic energy due to a single particle collision.
We consider three different models for the coefficient of restitution: (i) $\varepsilon$=const.,  (ii) a stepwise dependence of $\varepsilon$ on the impact velocity, which mimics the basic property of this coefficient, and (iii) a realistic model for $\varepsilon$ as a function of the impact velocity, derived from the contact mechanics of viscoelastic bodies. For the simplest model (i), both theoretical methods  yield the same coefficient of diffusion. For the more realistic models (ii) and (iii), however, the Green-Kubo approach is either restrictive or too complicated for practical application. Therefore we conclude that the Chapman-Enskog method  is preferable for deriving kinetic coefficients of granular gases.
}  

\section{Introduction}

A homogeneously initialized force free granular gas, i.e., a rarefied systems of inelastically colliding macroscopic particles, stays homogeneous during the first stage of its evolution while its temperature, that is the average kinetic energy of particles,  decays due to dissipative collisions. This state of a granular gas is called the {\em homogeneous cooling state}. At later stages, granular gases develop density inhomogeneities \cite{GoldhirschZanetti:1993,McNamara:1993} and pronounced spatial correlations of the velocity field \cite{BritoErnst:1998}. Throughout this article we consider granular gases in the homogeneous cooling state. 

Similar as ordinary molecular gases, granular gases in the homogeneous cooling state are characterized by their temperature and their velocity distribution function, which is close to the Maxwell distribution.  Like in molecular gases, impurities (guest particles), that is particles which differ from the gas particles by mass, size or other parameters, move irregularly due to random collisions with the surrounding gas particles. This {\em Brownian motion} of guest particles, i.e. diffusion, leads to uniform spreading of an initially localized ensemble of impurities over the available volume. If the guest particles are mechanically identical to the host particles, but may be somehow distinguished (e.g. by color), the process is called self-diffusion and the guest particles are called tracers. 

The processes of diffusion and self-diffusion in granular gases differ from those in molecular gases because of the dissipative nature of particle collisions. A collisions of dissipatively colliding particles is characterized by the coefficient of restitution $\varepsilon$. This coefficient relates the post-collision velocities $\vec{v}^{\,\prime}_{i}$ and $\vec{v}^{\,\prime}_{j}$ of the colliding particles $i$ and $j$ to their pre-collision velocities $\vec{v}_{i}$ and $\vec{v}_{j}$. For particles of identical mass the collision law reads
\begin{equation}
\label{eq:bincoll}
\begin{split}
  \vec{v}^{\,\prime}_{i}&=\vec{v}_i-\frac{1+\varepsilon}{2} \left(\vec{v}_{ij} \cdot \vec{e}\,  \right) \vec{e} \qquad \\
\vec{v}^{\,\prime}_{j}&=\vec{v}_j+\frac{1+\varepsilon}{2} \left(\vec{v}_{ij} \cdot \vec{e}\,  \right) \vec{e} \, , 
\end{split}
\end{equation}
where $\vec{v}_{ij}\equiv\vec{v}_i-\vec{v}_j$. The unit vector $\vec{e}\equiv \vec{r}_{ij}/r_{ij}$ with $\vec{r}_{ij}\equiv\vec{r}_i-\vec{r}_j$ describes the relative particle position  at the instant of the collision. For particles of different mass the corresponding relation has a similar form \cite{BrilliantovPoeschelOUP}. The collision law \eqref{eq:bincoll} takes only the normal component of the particles' relative velocity into account. Correspondingly, $\varepsilon$, is also called the {\em normal} coefficient of restitution -- the tangential component of the relative velocity is not affected by a collision (see \cite{BrilliantovPoeschelOUP}, Chap. 3.4 for a detailed discussion of this non-trivial approximation). In what follows we consider the normal coefficient of restitution.

Granular gases are never in thermodynamic equilibrium because of the permanent loss of energy due to dissipative particle collisions. Therefore, the basic law of equilibrium statistical mechanics, the equipartition of kinetic energy, is not applicable for granular gases. Consequently, the temperature of the gas particles may differ from the temperature of the guest particles \cite{GarzoMontanero:2004,BarratTrizac_GM:2002,Marconi1Puglisi:2002,BenNaimKrapivsky_EPJ:2002,ClellandHrenya:2002,Dahletal:2002,WangJinMa:2003,FeitosaMenon:2002,WildmanParker:2002,SantosDufty_PRL:2001,SantosDufty_PRE:2001}. Hence, in granular gases the temperature is species dependent. The violation of equipartition leads to many interesting phenomena, like criticality of the  Brownian motion \cite{SantosDufty_PRL:2001,SantosDufty_PRE:2001} and turns the analysis technically complicated. Here we restrict ourselves to the case of self-diffusion, when  the gas may be characterized by a single temperature, that is, we discuss the diffusion of tracers. In spite of being much simpler, self-diffusion  demonstrates the most salient properties of transport in granular gases. It allows to illustrate the application of the main theoretical approaches for the computation of the transport coefficients, the Green-Kubo and Chapman-Enskog approaches, without much technical complications. 

We show that for the case of a constant coefficient of restitution, the Green-Kubo approach, which has been initially elaborated for equilibrium systems, may be efficiently exploited. For a direct and thus simple application of this method, the dissipative gas dynamics can be mapped onto the corresponding dynamics of a system which remains in an effective steady-state. We will show that in this case the Green-Kubo and Chapman-Enskog methods yield the same result for the self-diffusion coefficient. For the case $\varepsilon=$const. such a simple map exists. 

The assumption of a constant coefficient of restitution simplifies the analysis drastically, however, it does not agree   with experimental observations \cite{BridgesHatzesLin:1984,KuwabaraKono:1987,McNamaraFalcon:2003}. Moreover, the assumption $\varepsilon=$const. contradicts the mechanics of materials \cite{Tanaka,LudingClementBlumenRajchenbachDuran:1994,LudingClementRajchenbachDuran:1996,BrilliantovSpahnHertzschPoeschel:1994,SchwagerPoeschel:1998,MorgadoOppenheim:1997} and even violates a dimension analysis \cite{Ramirez:1999}. Instead, experiments as well as theory show that the coefficient of restitution increases with decreasing impact velocity. Thus, for $g\to 0$ particles tend to collide elastically, that is $\varepsilon\to 1$ if surface effects are neglected.

This basic property motivates a drastically simplified collision model \cite{PoeschelBrilliantovSchwager:2003,NieBenNaimChen2002}:
\begin{equation}
  \label{eq:epsstep}
  \varepsilon(g) = \begin{cases}
    \varepsilon^* & \text{for~~~}  g>g^*\\
    1             & \text{for~~~}  g \le g^*\,,     
  \end{cases}
\end{equation}
where $g\equiv\vec{v}_{ij} \cdot \vec{e}$ is the normal component of the relative velocity $\vec{v}_{ij}$ at the impact. Thus, it is assumed that impacts at large normal relative velocity  occur with a constant coefficient of restitution $\varepsilon^*$, while collisions at small $g$ are elastic, i.e., the threshold $g^*$ separates elastic from inelastic collisions. The model parameters, $\varepsilon^*$ and $g^*$, are neither related to the material properties of particles nor derived from first principles of the collision dynamics. 

The coefficient of restitution as a function of the impact velocity may be also derived analytically by integrating Newton's equation of motion for a colliding pair, provided the particle interaction force is known. The simplest physically realistic mechanical model of dissipative collisions assumes  that particles interact by viscoelastic forces \cite{KuwabaraKono:1987,BrilliantovSpahnHertzschPoeschel:1994,MorgadoOppenheim:1997,Ramirez:1999}. If inelasticity is small, the solution of the contact problem for viscoelastic spheres yields the coefficient of restitution \cite{SchwagerPoeschel:1998,Ramirez:1999}
\begin{equation}
  \label{eq:epsvisco}
  \varepsilon_v(g)= 1-\gamma g^{1/5}+ \frac35 \gamma^2  g^{2/5}\mp\dots\,,
\end{equation}
where the coefficient $\gamma$ is a known function of the elastic and dissipative constants, i.e. of the Young modulus and the Poisson ratio of the particles material, and of the masses and radii of the particles (see \cite{SchwagerPoeschel:1998,Ramirez:1999} for details). 
 
We will show that for the case of impact-velocity dependent $\varepsilon$ there is no simple map of the dynamics of the dissipative system onto a corresponding dynamics of a steady-state system. This makes the application of the Green-Kubo method problematic. Therefore, for the case of an impact-velocity dependent coefficient of restitution the Chapman-Enskog approach is used. We apply the latter method to calculate the self-diffusion coefficient for the case of a simplified impact-velocity dependent coefficient of restitution as well as for realistic gases of viscoelastic particles. 

The rest of the paper is organized as follows: In Sec. II we
discuss the microscopic interpretation of the macroscopic diffusion
coefficient which suggests a route of its evaluation.
The generalization of the diffusion coefficient to
non-equilibrium systems, such as granular gases is also
discussed. In Sec. III, the formalism of the
pseudo-Liouville operator is addressed, as used for hard-sphere systems, and Sec. IV illustrates its application to granular gases in the homogeneous cooling state. Three different collision models introduced above are there addressed. The Green-Kubo approach is discussed in Secs. V and VI, while the Chapman-Enskog method is described in Secs. VII and VIII. In Secs. V, VI and VII we explain the general concept of these methods for the simple case $\varepsilon= {\rm const.}$ Then we consider the more general case of impact-velocity dependent coefficient of restitution. We summarize our findings in Sec. IX and in Appendixes A and B some technical detail are given.

\section{Macroscopic and microscopic meaning of the diffusion coefficient}

Consider a uniform gas of hard spheres of mass $m$ and diameter $\sigma$ whose particles collide due to the collision law Eq. \eqref{eq:bincoll}, specified by the coefficient of restitution $\varepsilon$. Let tracers particles be embedded in the gas and
let $f(\vec{r}, \vec{v},t)$ and $f_s(\vec{v}_s, \vec{r}_s,t)$ be respectively the one-particle distribution function of the gas particles and of the tracers. Then the local densities are defined by
\begin{equation}
\label{eq:def_n0n}
\begin{split}
  n_s(\vec{r}_s,t)&=\int d \vec{v}_s f_s(\vec{v}_s, \vec{r}_s,t) \\
  n(\vec{r},t)&=\int d \vec{v} f (\vec{v}, \vec{r},t) \,.
  \end{split}
\end{equation} 
If the concentration of the tracers is not uniform, a diffusion flux of these particles arises, which is directed opposite to the concentration gradient:
\begin{equation}
\label{eq:Js_nablans}
\vec{J}_s \left(\vec{r}\,\right)= - D \vec{\nabla} n_s \left(\vec{r}\,\right) \,.
\end{equation}
 With the continuity equation 
\begin{equation}
\frac{\partial n_s \left(\vec{r}\,\right) }{\partial t} + \vec{\nabla} \vec{J}_s \left(\vec{r}\,\right)= 0
\end{equation}
we obtain the diffusion equation
\begin{equation}
\label{difcanon}
\frac{\partial n_s \left(\vec{r}\,\right) }{\partial t} = D \vec{\nabla}^2 n_s \left(\vec{r}\,\right)\,.
\end{equation} 
The diffusion coefficient $D$ in  Eq. \eqref{eq:Js_nablans}, thus, relates the macroscopic flux of particles $\vec{J}_s \left(\vec{r}\,\right)$ to the macroscopic concentration gradient $ \vec{\nabla} n_s \left(\vec{r}\,\right)$. 

However, there is also a microscopic interpretation of the diffusion coefficient: It defines the time dependence of the mean square displacement of a tracer particle. Assume at $t=0$ there is an ensemble of impurities located at the origin, $\vec{r}=0$. Then the mean square displacement of the tracers at time $t$ reads  
\begin{equation}
\label{sqavdef}
\left< \left[ \vec{r}_s (t) \right]^2  \right>\equiv 
\frac{1}{N_s} \int d \vec{r} \, \left[ \vec{r}\,(t)\right]^2 n_s \left(\vec{r},t\right) \,,  
\end{equation} 
where $N_s=\int d \vec{r} \, n_s \left(\vec{r},t\right)=\text{const.}$ is the total number of tracers. Now we multiply both sides of Eq. \eqref{difcanon} by $\vec{r}^{\,2}/N_s$ and integrate over $d\vec{r}$:
\begin{equation}
\frac{\partial}{\partial t} \frac{1}{N_s} \int d \vec{r} \, r^2 n_s = D \, \frac{1}{N_s}   \int d \vec{r} \, r^2 \vec{\nabla}^2 n_s\,,   
\end{equation}
which turns by means of Eq. \eqref{sqavdef} and Green's theorem \cite{AbramowitzStegun:1965} for the right-hand side into
\begin{equation}
\frac{d }{d t} \left< \left[ \vec{r}_s (t) \right]^2  \right> = 
D \, \frac{1}{N_s} \int d \vec{r} \, n_s \vec{\nabla}^2  r^2 =6 D\,,
\end{equation} 
that is, we obtain a relation between the mean square displacement of the tracers and the coefficient of diffusion:
\begin{equation}
\label{sqavdef1}
\left< \left[ \vec{r}_s \, (t) \right]^2  \right> =6Dt \, . 
\end{equation} 

This equation relates the diffusion coefficient $D$, i.e. a macroscopic quantity, to a microscopic quantity, namely the mean square displacement. Therefore, Eq. \eqref{sqavdef1} suggests a method to compute the diffusion coefficient by evaluating $\left< \left[ \vec{r}_s \, (t) \right]^2  \right>$.
To find this quantity we apply the kinematic relation 
\begin{equation}
\vec{r}_s \, (t)= \int_0^t \vec{v}_s \left( t_1 \right) dt_1~~~~~ \text{(}\vec{r}_s=0~~\text{at}~~t=0\text{)}   
\end{equation}
and write 
\begin{equation}
\left< \left[ \vec{r}_s \, (t) \right]^2  \right>  =\left< \int_0^{t}\vec{v}_s\left(t_1\right)dt_1 \cdot \int_0^t  \vec{v}_s\left(t_2\right) dt_2  \right> \, . 
\label{delR}
\end{equation}
Using the symmetry of the velocity autocorrelation function, 
\begin{equation}
\label{defVAF}
K_v\left(t_1, t_2\right)\equiv \left< \vec{v}_s \left(t_1\right) \cdot \vec{v}_s\left(t_2\right) \right> =K_v\left(t_2, t_1\right) \, , 
\end{equation}  
we can recast Eq. \eqref{delR} into the form
\begin{equation}
\label{drviaKvgen}
\left< \left[ \vec{r}_s \, (t) \right]^2 \right>  = 2 \int_0^{t} d t_1 \int_{t_1}^{t} d t_2 K_v \left( t_2, t_1 \right)\,.
\end{equation}  
If the system is in a steady state its properties are invariant under a shift of time, thus, the velocity autocorrelation function depends only on the time lag $\tau \equiv t_2-t_1$, that is, $K_v\left( t_2, t_1\right)=K_v(\tau)$ and decays with a characteristic relaxation time $\tau_v$. Transforming the variables   $(t_1,\, t_2) \to (\tau, \, \tau_1)$ with $\tau_1=t_1$ the mean square displacement reads 
\begin{equation}
\label{2drviaKvgen}
\left< \left[ \vec{r}_s \, (t) \right]^2  \right> = 2t \int_0^{t} d \tau
\left< \vec{v}_s \, (0) \cdot \vec{v}_s \, (\tau) \right> \left( 1- \frac{\tau}{t} \right)\,.
\end{equation}  
For large time as compared with the characteristic time, i.e.\ for $t \gg \tau_v$, we obtain
\begin{equation}
\label{3drviaKvgen}
\left< \left[ \vec{r}_s \, (t) \right]^2  \right> = 2t \int_0^{t} d \tau
\left< \vec{v}_s \, (0) \cdot \vec{v}_s \, (\tau) \right> = 6D t
\end{equation}  
and consequently the diffusion coefficient
\begin{equation}
\label{Dvel}
D=\frac13 \, \int_0^{\infty}\,\left< \vec{v}_s \, (0) \cdot \vec{v}_s \, (t)\,\right> dt \,.
\end{equation}  
Equation \eqref{Dvel} expresses the macroscopic transport coefficient $D$ by the time integral of the microscopic velocity correlation function. This fluctuation-dissipation relation expresses a dissipative quantity (diffusion coefficient) via  a fluctuating quantity (particle velocity). Similar expressions relate other transport coefficients such as viscosity, thermal conductivity, etc. to the corresponding fluctuating quantities \cite{resibua}. The method of calculation of the transport coefficients via the microscopic time-correlation functions is called Green-Kubo approach. Initially, the Green-Kubo theory has been developed for equilibrium systems of elastic particles (e.g. \cite{resibua}). Later it has been  generalized for dissipative gases \cite{GoldhirschNoije:2000,BreyDuftyRuizMontero:2003,BreyRuizMonteroPRE:2004}. For the case of self-diffusion the velocity correlation function in Eqs. (\ref{delR}-\ref{Dvel}) for the tracer particles is equal to that of the surrounding gas particles. 

The derivation given above applies to equilibrium systems or systems in a steady state. Granular gases are, however, never in a steady state due to persistent cooling. Therefore, the concept of the diffusion coefficient has to be generalized to non-equilibrium systems and a time-dependent diffusion coefficient $D(t)$ (also called diffusivity) must be used \cite{EsipovPoeschel:1995,BrilliantovPoeschel:1998d}:
\begin{equation}
\label{eq:diffusivity}
\left< \left[ \vec{r} \, (t) \right]^2  \right>  = \int^t D(t^{\prime}) d t^{\prime} \,.
\end{equation}  
As we show below, in certain cases, the dynamics of granular gases may be mapped onto the dynamics of a steady-state system. In this case the above Green-Kubo derivation of the diffusion coefficient is still applicable. 

\section{Dynamics of dissipative hard sphere systems: General approach}

Because of the singular character of the hard-sphere interaction (the interaction potential jumps to infinity at the point of particles contact) it is not possible to apply directly Hamilton's equations to describe the dynamics of hard-sphere systems. Nevertheless, the dynamics is still completely deterministic. Indeed, between collisions the particles move along straight lines at constant velocities, and the post-collision velocities are uniquely determined by the collision law Eq. \eqref{eq:bincoll}. This allows to construct the pseudo-Liouville operator ${\cal L}$ which governs the dynamics of the system. Let $A(t)$ be an arbitrary function of particle coordinates and velocities, then the evolution equation for this quantity reads
\begin{equation}
\label{L}
\frac{d}{dt}  A\left(\vec{r}, \vec{v}\, \right) = {\cal L} A\left(\vec{r}, \vec{v}\, \right) \,,
\end{equation}  
where the pseudo-Liouville operator has the form (e.g.  
\cite{DuftyBreyLutsko:2002}) 
\begin{equation}
\label{eq:defL}
{\cal L}= \sum_{i=1}^{N} \vec{v}_i \cdot \vec{\nabla}_i + \sum_{i=1}^{N}\sum_{j > i}^{N}\hat{T}_{ij}\,
\end{equation}  
with the binary collision operator 
\begin{equation}
\label{eq:Tij}
\hat{T}_{ij}=\sigma^{2} \int d\vec{e}\, \Theta\left(- \vec{v}_{ij} \cdot \vec{e} \, \right) \left|\vec{v}_{ij} \cdot \vec{e} \, \right| \delta \left( \vec{r}_{ij}- \sigma \vec{e} \, \right)\left[\hat{b}_{ij}^{\vec{e}}-1 \right] \, . 
\end{equation}  
Here $N$ is the number of particles, $\vec{r}_i$ and $\vec{v}_i$ are  respectively the position and velocity of particle $i$, $\vec{r}_{ij}\equiv\vec{r}_i-\vec{r}_j$, $\vec{v}_{ij}\equiv\vec{v}_i-\vec{v}_j$ and $\vec{e}\equiv\vec{r}_{ij}/r_{ij}$. The factor $\left|\vec{v}_{ij} \cdot \vec{e} \, \right|$ in the integrand of Eq. \eqref{eq:Tij} gives the length of the collision cylinder, the Heaviside step function $\Theta\left(- \vec{v}_{ij} \cdot \vec{e} \, \right)$ selects approaching particles and the  $\delta $ function specifies the unit vector $\vec{e}$ at the instant of the collision. The operator $\hat{b}_{ij}^{{\vec{e}}}$ is defined by
\begin{equation}
\label{eq:defofbe}
\hat{b}_{ij}^{\vec{e}} F \left(\vec{r}_{i},  \vec{r}_{j}, \vec{v}_{i},\vec{v}_{j} \cdots \right)=F \left(\vec{r}_{i}, \vec{r}_{j},  \vec{v}^{\,\prime}_{i},\vec{v}^{\,\prime}_{j} \cdots \right) \,,
\end{equation}
where $F$ is some function of the positions and coordinates. The post-collision velocities of the colliding pair, $\vec{v}^{\,\prime}_{i}$ and $\vec{v}^{\,\prime}_{j}$, are given in terms of their pre-collision values according to Eq. \eqref{eq:bincoll}. The pseudo-Liouville operator defined above describes the evolution of dynamical variables, such as positions of particles, their velocities, etc. It has been introduced first to describe the evolution of a system of {\em elastic} hard spheres \cite{ErnstEtAl:1969,ErnstDorfman:1972}.

The velocity time correlation function may be found by performing the formal integration of Eq. \eqref{L} for the variable $A=\vec{v}$ for $t>t^{\prime}$:
\begin{equation}
\vec{v} \left( t \right)= e^{ {\mathcal L} \left(t-t^{\,\prime} \right)} \vec{v} \left( t^{\,\prime}\right) \,,
\label{evolA}
\end{equation}
yielding
\begin{equation}
\left< \vec{v} \left(t^{\prime}\right) \vec{v} \left(t\right) \right >=\int d\Gamma \rho\left(t^{\prime}\right) \vec{v} \left(t^{\prime}\right) e^{ {\mathcal L} \left(t-t^{\prime}\right)} \vec{v} \left(t^{\prime}\right)\,,
\label{evolAA}
\end{equation}
where $\int d\Gamma$ denotes integration over all degrees of freedom and  $\rho\left(t \right)=\rho\left(\vec{r}_1, \ldots, \vec{r}_N, \vec{v}_1, \ldots, \vec{v}_N, t \right)$ is the $N$-particle distribution function which depends on temperature $T$, particle number density $n$, etc. Although  Eq. \eqref{evolAA} provides  the velocity correlation function $K_v$, whose time integral yields the diffusion coefficient, it may not be easily used in this general form. Instead, for practical application of the  Green-Kubo method, in general, further approximations are required. 

\section{Homogeneous cooling state} 

\subsection{General properties} 

We consider a granular gas in the homogeneous cooling state when the number density of the gas $n=N/V$ is uniform in the gas volume $V$, while the gas cools down due to inelastic particle collisions. 

The evolution of the gas temperature may be easily obtained from its definition as the mean kinetic energy of the particles, 
\begin{equation}
\label{eq:def_T}
\frac32 n \, T(t) = n  \left< \frac12 m v^2 \right>_t=\int \frac{mv^2}{2} f(\vec{v},t) d\vec{v} \,. 
\end{equation}
From the definition of the Liouville operator, Eqs. \eqtoeqref{eq:defL}{eq:Tij} follows
\begin{equation}
\label{eq:dT_T0_dt}
\frac32\, \frac{dT}{dt} =\frac{m}{2} \, \frac{d}{dt} \left< v_1^2 \right>_t= \frac{m}{2}\,  \left<  {\mathcal L} \,v_1^2 \right>_t \,, 
\end{equation}
where  $\left< \ldots \right>_t$ denotes averaging using the $N$-particle distribution function $\rho(t)$ at time $t$. With Eqs. \eqtoeqref{eq:defL}{eq:Tij} and exploiting the identity of the particles we obtain
\begin{equation}
\label{eq:dT_T0_dt1}
\frac32\, \frac{dT}{dt} = \frac{m}{2}  \left<  \sum_{i<j} \hat{T}_{ij}  v_1^2 \right>_t = 
\left(N-1\right) \frac{m}{2} \, \left< \hat{T}_{12} v_1^2 \right>_t  \,.
\end{equation}
Using the definition of the binary collision operator, Eq. \eqref{eq:Tij}, and 
\begin{equation}
 \left(\hat{b}_{ij}^{\vec{e}}-1 \right) v_1^2=\left(v^\prime_1\right)^2-v_1^2 
\end{equation}
Eq. \eqref{eq:dT_T0_dt1} turns into
\begin{multline}
\label{TviaTij1}
       \frac32\, \frac{dT}{dt} =\frac{m}{2} \,\left( N-1 \right) \int d\vec{r}_1 \ldots d\vec{r}_N \int d\vec{v}_1 \ldots d\vec{v}_N \rho(t) \times \\
       \sigma^2  \int d\vec{e} \, \Theta\left(-\vec{v}_{12} \cdot \vec{e} \, \right) \left| \vec{v}_{12} \cdot \vec{e} \, \right| \delta\left(\vec{r}_{12}-\sigma \vec{e} \, \right)
       \left[ \left(v^\prime_1\right)^2-v_1^2 \right] \, .
\end{multline}
Using the definition of the two-particle correlation function \cite{resibua} and the hypothesis of molecular chaos, for a uniform system we may further write  
\begin{equation}
\label{deff2}
\begin{split}
N(N-1)&\int d\vec{r}_3 \ldots d\vec{r}_N \int d\vec{v}_3 \ldots d\vec{v}_N \rho(t)\\
&\equiv
f_2\left(\vec{r}_1,\vec{r}_2, \vec{v}_1, \vec{v}_2,t\right) \, \\[0.2cm]
&=g_2\left(r_{12} \right)f\left( \vec{v}_1,t\right)f\left(\vec{v}_2,t\right) \, , 
\end{split}
\end{equation} 
where $g_2\left(r_{12}\right)$ is the pair correlation function. Substituting finally the latter expression into Eq. \eqref{TviaTij1} yields 
\begin{multline}
\label{eq:dTdt1}
\frac32\, \frac{dT}{dt} = \frac{m}{2n}  g_2\left(\sigma \right) \sigma^2 \int d\vec{v}_1 d\vec{v}_2  \int d\vec{e}  \, \Theta\left(-\vec{v}_{12} \cdot \vec{e} \, \right)\\ 
 \times \left| \vec{v}_{12} \cdot \vec{e} \, \right|  f\left( \vec{v}_1,t\right)f\left(\vec{v}_2,t\right) \left[ \left(v^\prime_1\right)^2-v_1^2 \right]  \,.
\end{multline}
Introducing the time-dependent thermal velocity, the reduced velocities, and the scaled velocity distribution function defined respectively by
\begin{equation}
\label{eq:defVDFscal}
  v_T(t)\equiv\sqrt{\frac{2T}{m}}\,,~~~\vec{c}_i\equiv\frac{\vec{v}_i}{v_T}\,,~~~f\left(\vec{v}, t\right)=\frac{n}{v_T^3}\tilde{f}\left(\vec{c}, t\right)
\end{equation}
we write the last equation in a more traditional form (e.g. \cite{BrilliantovPoeschelOUP}),   
\begin{eqnarray}
\label{eq:dTdtcanon}
\frac{dT}{dt} &=& -\zeta T \\
\label{eq:zetadef}
 \zeta        &=& \frac23 n \sigma^2 g_2\left(\sigma \right) v_T \mu_2 \\
\label{eq:mu2def}
 \mu_2 &=& -\frac12 \int d\vec{c}_1 d\vec{c}_2  \int d\vec{e} \, \Theta\left(-\vec{c}_{12} \cdot \vec{e} \, \right) \\
&&   \times \left| \vec{c}_{12} \cdot \vec{e} \, \right|
\tilde{f}\left(
\vec{c}_1,t\right)\tilde{f}\left(\vec{c}_2,t\right) \Delta
\left(c_1^2+c_2^2\right)  \nonumber \, ,
\end{eqnarray}
where the symmetry with respect to the exchange of particle indices $1 \leftrightarrow 2$ has been exploited. The notation
\begin{equation}
\Delta \left(\psi \right)\equiv \psi^{\prime} - \psi
\end{equation}
describes the change of a dynamical variable $\psi$ due to a collision, with $\psi^{\prime}$ being the after-collision value. The contact value of the pair correlation function reads \cite{CarnahanStarling} 
\begin{equation}
 g_2\left(\sigma \right)= \frac{1-\frac12 \eta}{(1-\eta)^3}\,,~~~~~ \text{with} ~~ \eta=\frac16 \pi n \sigma^3  \,.
\end{equation}
For a granular gas in the homogeneous cooling state the reduced distribution function $\tilde{f}\left(\vec{c}, t\right)$ is close to the Maxwellian \cite{EsipovPoeschel:1995,GoldshteinShapiro1:1995,NoijeErnst:1998,PoeschelBrilliantovLNP:2003,BrilliantovPoeschelOUP}. It may be written with good accuracy, keeping only the first two non-vanishing terms of the Sonine polynomial expansion \cite{GoldshteinShapiro1:1995,NoijeErnst:1998,PoeschelBrilliantovLNP:2003,BrilliantovPoeschelOUP}:
\begin{equation}
\label{eq:VDF_HCS}
\tilde{f}\left(\vec{c}\,, t \right)= \phi(c) \left[ 1+ a_2 (t) S_2\left(c^2\right) \right]  \, , 
\end{equation}
where 
\begin{equation}
\label{eq:Maxdistr}
\phi(c) = \pi^{-3/2} \exp\left(-c^2\right)  
\end{equation}
is the rescaled Maxwell distribution and
\begin{equation}
\label{eq:S2defin}
 S_2(c^2) = \frac12 c^4 -\frac52 c^2 + \frac{15}{8} \,
\end{equation}
is the second-order Sonine polynomial.
The second Sonine coefficient $a_2$ characterizes the shape of the velocity distribution function. It sensitively depends on the particular collision model, which we discuss separately below. 

According to Eq. \eqref{eq:zetadef}, to find the cooling coefficient $\zeta$ we need to compute $\mu_2$, which may be written as an integral of the general form
\begin{equation}
  \label{eq:AppType}
  \int\!\! d\vec{c}_1\! \int\!\! d\vec{c}_2\! \int\!\! d\vec{e}\, \Theta\left(-\vec{c}_{12} \cdot \vec{e}\,\right) \left|\vec{c}_{12} \cdot \vec{e}\,\right|\phi(c_1) \phi(c_2) {\rm Expr}\left(\dots\right) \,,
\end{equation}
where 
\begin{equation}
\text{Expr}=-\frac12 \Delta \left( c_1^2 +c_2^2 \right)  \left[ 1+ a_2 S_2\left(c_1^2\right) \right] \left[ 1+ a_2 S_2\left(c_2^2\right) \right] \, . 
\end{equation}
We call integrals of the type Eq. \eqref{eq:AppType} {\em kinetic integrals} \cite{PoeschelBrilliantovLNP:2003,BrilliantovPoeschelOUP}. The kernel ${\rm Expr}\left(\dots\right)$ may contain the pre- and post-collision velocities, $\vec{c}_{1/2}$, $\vec{c}^{\,\prime}_{1/2}$, the Sonine polynomials $S_i$ of these velocities, the Sonine coefficients $a_i$, the unit vector $\vec{e}$, the relative velocity $\vec{c}_{12}$, the center of mass velocity of the colliding particles, $\vec{C}$, scalar products of the mentioned vectors, and other variables. Further, the coefficient of restitution may enter the kernel ${\rm Expr}\left(\dots\right)$, either $\varepsilon=\text{const}$ or as a function of the impact velocity $\varepsilon=\varepsilon(g)$ as discussed below. 

The methods for computing kinetic integrals are explained in detail in \cite{PoeschelBrilliantovLNP:2003,BrilliantovPoeschelOUP}. Kinetic integrals may be represented by sums of standard-type integrals which allow for a solution by means of computational symbolic algebra. Many physical quantities of interest may be expressed in terms of kinetic integrals \cite{PoeschelBrilliantovLNP:2003,BrilliantovPoeschelOUP}. Let us consider in more detail the collision models discussed in the Introduction. 

\subsection{Granular gases with ${\mathbf{\varepsilon}}$\lowercase{=const}}

As mentioned above, the coefficient of restitution is a function of the impact velocity. Nevertheless, here we assume $\varepsilon=$const. to simplify the explanation of the basic approaches. For this case the second Sonine coefficient reads \cite{NoijeErnst:1998,BrilliantovPoeschelStability:2000}  (see Appendix \ref{app:sonine}, where the derivation of $a_2$ is outlined):
\begin{equation}
\label{eq:a2Ctn}
a_2 = \frac{16 (1-\varepsilon)(1-2 \varepsilon^2)}{81 -17 \varepsilon +30 \varepsilon^2 (1-\varepsilon)} \, .
\end{equation}
Equation \eqref{eq:a2Ctn} gives the second Sonine coefficient in a linear approximation with respect to $a_2$, the next-order solution is derived in \cite{BrilliantovPoeschelStability:2000}. Note that for $\varepsilon$=const. the shape of the scaled distribution function $\tilde{f}$ and, hence, the Sonine coefficient do not depend on 
time.\footnote{MD simulations of two-dimensional granular gases show that the second Sonine coefficient $a_2$ relaxes to the theoretical value within a few collisions per particle \cite{HuthmannOrzaBrito:2000,Nakanishi:2003}. At later times, however, $a_2$ deviates from the theoretical result due to accumulating correlations between the particles \cite{Nakanishi:2003}, see Appendix \ref{app:sonine}.}

The coefficient $\mu_2$ may be found by means of the computation method elaborated for the kinetic integrals \cite{BrilliantovPoeschelStability:2000}:
\begin{equation}
\label{eq:mu2}
\mu_2=\sqrt{2 \pi} \left(1-\varepsilon^2 \right) \left( 1+ \frac{3}{32}a_2 \right)^2 \,.
\end{equation}
Substituting $\mu_2$ into Eq. \eqref{eq:zetadef} yields the cooling coefficient $\zeta$. Since $\mu_2$ does not depend on temperature for $\varepsilon = {\rm const.}$, the cooling coefficient scales as $\zeta \propto T^{1/2}$, see Eqs. (\ref{eq:defVDFscal},\ref{eq:zetadef}). Hence, from Eqs. (\ref{eq:dTdtcanon}, \ref{eq:zetadef}) we obtain for the evolution of temperature 
\begin{equation}
\label{eq:TdecayEpsCon} T=\frac{T(0)}{(1+t/\tau_0)^2}\, ; \qquad
\tau_0^{-1}=\frac{\mu_2}{3 \sqrt{8 \pi}} \tau_c(0)^{-1} \,,
\end{equation}
where $\mu_2$ and $\tau_c$ are given respectively by
Eq. \eqref{eq:mu2} and Eq. \eqref{tauc01}.

\subsection{Granular gases with stepwise impact-velocity dependent coefficient of restitution}
 
For granular gases of particles which collide with the simplified stepwise impact-velocity dependent coefficient due to Eq. \eqref{eq:epsstep}, the velocity distribution function is close to a Maxwell distribution with good accuracy \cite{PoeschelBrilliantovSchwager:2003}. Therefore, $a_2 \approx 0$ for these gases, and the coefficient $\mu_2$ reads (see Appendix \ref{app:zeta})
\begin{equation}
\label{eq:mu2_fin}
\mu_2= \sqrt{2 \pi}  \left( 1-\varepsilon^{* \, 2} \right) \left(1+ \frac{m g^{*\, 2}}{4T} \right) 
\exp \left( - \frac{mg^{*\, 2}}{4T} \right) \,.
\end{equation}
Substituting  \eqref{eq:mu2_fin} into Eq. \eqref{eq:zetadef} yields the cooling coefficient $\zeta$ for this collision model.
According to Eqs. \eqref{eq:mu2_fin} and \eqref{eq:zetadef}, $\zeta$ decays exponentially fast as a function of temperature, $\zeta \propto \exp(-{\rm const.} /T)$, that is, the cooling of the gas slows down exponentially as the gas evolves. Asymptotically, for $t \to\infty$, this leads to a logarithmically slow decay of the temperature 
\cite{PoeschelBrilliantovSchwager:2003}:
\begin{equation}
  \label{eq:TepsStep}
  T(t)=\frac{g_0^2}{2}\frac{T_0}{\log\alpha t}\,,~~~\alpha=2g_0\left(1-\varepsilon^{*2}\right)g_2(\sigma)\sigma^2 n\sqrt{\frac{\pi T_0}{2m}}\,,
\end{equation}
where $g_0=g^*/v_T(0)$.

\subsection{Granular gases of viscoelastic particles}

For gases of viscoelastic particles the shape of the velocity distribution function and, thus, the Sonine coefficients depend on time. The time scale of diffusion processes -- the hydrodynamic time scale -- is much larger than the mean collision time. In this case one obtains approximate expressions \cite{BrilliantovPoeschel:2003}:
\begin{equation}
\label{eq:a2viscoadiab}
a_2  = a_{21} \delta^{\, \prime}  + a_{22} \delta^{\, \prime \, 2}  + \dots \, ,
\end{equation}
where $a_{21}  \approx -0.388$\footnote{The numerical values of
the constants $a_{21}$, $a_{22}$, $C_1$, $\omega_0$, $\omega_2$,
and $q_0$  are known analytically
\cite{BrilliantovPoeschel:2003,BrilliantovPoeschel:2000visc,BrilliantovPoeschelOUP}.}
and  $a_{22} \approx 2.752$ are pure numbers  and $\delta^{\,
\prime}$ is the time-dependent dissipation coefficient,
\begin{equation}
\label{eq:defdelprime} 
\delta^{\, \prime} (t) = \frac{\gamma}{C_1} \left[ v_T(t) \right] ^{1/5}
\end{equation}
with $C_1 \approx 1.1534$. 

Similarly, solving the kinetic integral, the coefficient $\mu_2$ is obtained for gases of viscoelastic particles,
\begin{equation}
\label{eq:defmu2Visco}
\mu_2 = \omega_0 \delta^{\, \prime} - \omega_2 \delta^{\, \prime \, 2}+\dots \,,
\end{equation}
where $\omega_0 \approx 6.485$ and $\omega_2 \approx 9.888$.

By means of Eqs. \eqref{eq:zetadef}, \eqref{eq:defdelprime}, and \eqref{eq:defmu2Visco} we obtain in leading-order, $\zeta \sim T^{3/5}$ and find \cite{SchwagerPoeschel:1998,BrilliantovPoeschelOUP}
\begin{equation}
\label{eq:TdecayEpsVis} T=\frac{T(0)}{(1+t/\tau_0)^{5/3}}\, ;
\qquad \tau_0^{-1}=\frac{16}{5}q_0  \frac{\gamma}{C_1} \left(
\frac{T_0}{m} \right)^{1/10} \tau_c(0)^{-1} \,
\end{equation}
where $q_0 \simeq 0.173$. Hence different collision models predict
different evolution of temperature, which asymptotics ranges from
the power law, $T  \to  t^{-\alpha}$ with $\alpha=2$ for
$\varepsilon = {\rm const.}$ and $\alpha =5/3$ for viscoelastic
particles, to the logarithmic decay, $T \to (\log t)^{-1}$ for the
stepwise impact-velocity dependent $\varepsilon$.

\section{Velocity correlation function for $\mathbf{\varepsilon}$\lowercase{=const}}

Again we consider a granular gas in the homogeneous cooling state and assume $\varepsilon$=const. Let time be measured in units of the mean collision time $\tau_c(t)$,
\begin{equation}
d \tau = \frac{dt}{\tau_c(t)}\,, \quad\quad \tau(t) = \int_0^t \frac{dt^{\prime}}{\tau_c(t^{\prime})} \,  
\label{eq:tau_t_viatauc}
\end{equation} 
with\footnote{To be precise, $\tau_c$ differs slightly from the actual mean collision time due to deviations of the velocity distribution from the Maxwell distribution \cite{BrilliantovPoeschelOUP}.}  
\begin{equation}
\label{tauc01}
\tau_c^{-1}(t) = 4\sqrt{\pi} g_2(\sigma)\sigma^2 n \sqrt{\frac{T(t)}{m}} \,. 
\end{equation}
The dynamics of the system is stationary when replacing the true (laboratory) time by the rescaled time $\tau$. Neglecting particle-particle correlations (see \cite{BrilliantovPoeschelOUP}) the dynamics of the system equals the dynamics of an equilibrium system. Thus, by rescaling time, the  dynamics of the dissipative gas is mapped onto the stationary dynamics of a gas of elastically colliding particles \cite{BreyDuftyRuizMontero:2003,BreyRuizMonteroGarciaRojo:1999,BrilliantovPoeschelOUP}. In this state the collision frequency and temperature are constant and the only difference from an equilibrium molecular gas is that the collisions still follow the rules of the  dissipative impact, Eq. \eqref{eq:bincoll}, with the coefficient of restitution $\varepsilon$. The velocity time-correlation function may be then written in the form 
\begin{equation}
\left< \vec{v}_1 \left(t^{\prime}\right)\cdot  \vec{v}_1(t)\right> = v_T \left(t^{\prime}\right) v_T \left(t\right) \left< \vec{c}_1 \left(t^{\prime}\right)\cdot  \vec{c}_1(t)\right>\,,
\label{eq:vv_cc}
\end{equation} 
i.e., we need to find the time-correlation function for the reduced velocity $\vec{c}\,(t)$. In the new time scale the collision frequency and temperature are constant and the system is in a stationary steady state. Consequently, the time correlation function $\left< \vec{c}_1 \left(\tau^{\prime}\right)\cdot  \vec{c}_1(\tau)\right>$ does not depend  on the times $ \tau^{\prime}$ and $\tau$ separately, but only on  their  difference $|\tau^{\prime} -\tau|$. 

Adopting the molecular chaos hypothesis, the velocities of two colliding particles are not correlated. Hence, the collision process for a particle is stationary and Markovian. If we additionally assume that the process is Gaussian, we deduce from Doob's theorem \cite{resibua} an exponential velocity correlation function:
\begin{equation}
  \begin{split}
\left< \vec{c}_1 \left( \tau^{\prime} \right) \cdot  \vec{c}_1 \left( \tau \right) \right> &= \left< c_1^2 \right> 
\exp\left(-\frac{\left| \tau - \tau^{\prime} \right|}{\hat{ \tau }_v}\right) \\
&=\frac32 \exp\left(- \frac{\left| \tau- \tau^{\prime} \right|}{\hat{\tau}_v}\right) \, , 
  \end{split}
\label{eq:VCF_ctauctau}
\end{equation} 
with $\hat{\tau}_v$ being the velocity relaxation time, measured in units of the mean collision time $\tau_c(t)$. Obviously,  $\hat{\tau}_v$ is constant since the system is in a stationary state. 

To compute the time derivative of the correlation function at time $t^{\prime}=t$, we use the relations
\begin{eqnarray} 
  \frac{dv_T}{dt}&=& \frac{1}{2T} \, \dot{T} \, v_T(t) = - \frac12 \zeta(t)  v_T(t) \\
  \frac{3}{2} v_T^2(t)&=&  \left< v_1^2 (t)\right>_t
\end{eqnarray} 
which follow from the definitions of $v_T$ and $T$, and introduce the laboratory-time dependent velocity relaxation time
\begin{equation}
\label{eq:labTau}
\tau_v (t) = \hat{\tau}_v \tau_c (t)\,,   
\end{equation}
which corresponds to the reduced relaxation time $\hat{\tau}_v$. Eventually, we obtain
\begin{equation}
\frac{d}{dt}  \left. \left< \vec{v}_1 \left(t^{\prime}\right)\cdot  \vec{v}_1(t)\right> \right|_{t^{\prime}=t}   
  = - \left<  v_1^2(t) \right> \left[  \frac12 \zeta(t) + \tau_v^{-1} (t) \right]  \,. 
\label{eq:derVCF_ttprime}
\end{equation} 
On the other hand, using Eq. \eqref{evolAA}, we obtain for the same quantity 
\begin{equation}
  \begin{split}
    \frac{d}{dt} & \left. \left< \vec{v}_1 \left(t^{\prime}\right)\cdot  \vec{v}_1(t)\right> \right|_{t^{\prime}=t}\\ 
    &= \frac{d}{dt} \left. \int d\Gamma \rho\left(t^{\prime}\right) 
      \vec{v}_1 \cdot e^{ {\mathcal L} \left(t-t^{\prime}\right)} \vec{v}_1 \right|_{t^{\prime}=t}
    =\left< \vec{v}_1 \cdot  {\mathcal L } \vec{v}_1 \right>_t \\[0.2cm]
    &= (N-1) \left< \vec{v}_1 \cdot  \hat{T}_{12} \vec{v}_1 \right>_t  \, .  
             \end{split}
\label{eq:vv_rho_vLv}
\end{equation}
The last equality  in Eq. \eqref{eq:vv_rho_vLv} follows from the properties of the Liouville operator, Eq. \eqeqref{eq:defL}{eq:Tij} and from the identity of the particles. Comparing Eqs. \eqref{eq:vv_rho_vLv} and \eqref{eq:derVCF_ttprime} we obtain the velocity relaxation time $\tau_v (t)$:
\begin{equation}
  \begin{split}
    \tau_v^{-1}\left(t\right)&= - (N-1) \frac{\left< \vec{v}_1 \cdot \hat{T}_{12} \, \vec{v}_1\right>_{t}}{\left< \vec{v}_1^{\,2} \right>_{t}}- \frac12 \zeta(t) \\
&= \tau_{v,\,{\rm ad}}^{-1}(t) - \frac12 \zeta (t)\, .
\end{split}
\label{eq:vTv_zeta}
\end{equation} 
Physically, the first term on the right-hand side describes the relaxation (i.e. the decay) of the velocity correlation function due to collisions which randomize the directions and amplitudes of the particle velocities. It has the same nature as for molecular gases where particles collide elastically. The second terms describes the decay of the velocity correlation function due to the decrease of the thermal velocity of the cooling gas. For almost elastic particle collisions, $\varepsilon\lesssim 1$, each particle suffers a large number of collisions before the temperature decays noticeably. Thus, the velocity relaxation time is dominated by the first term and the decrease of the thermal velocity on the time scale of the velocity relaxation $\tau_v$ may be neglected, yielding the {\em adiabatic approximation of the velocity relaxation time}, $\tau_v \approx \tau_{v,\,{\rm ad}}$. 

Hence, Eqs. \eqref{eq:vv_cc}, \eqref{eq:VCF_ctauctau}, \eqref{eq:vTv_zeta}, and \eqref{eq:labTau} define the velocity correlation function of granular gases when $\varepsilon=$const is assumed. Note that in spite of the molecular chaos assumption, in the laboratory time the velocity correlation function does not decay exponentially as it does for gases of elastic particles. 

\section{Green-Kubo diffusion coefficient for $\mathbf{\varepsilon}$\lowercase{=const.}}

To find the diffusion coefficient it is convenient to use the reduced time, measured in collision units and the corresponding reduced length
\begin{equation}
\vec{r} = r_0 \hat{\vec{r}} \, , \quad\quad  r_0 \equiv v_T \left(t\right) \tau_c(t) \, .  
\label{eq:reduc_r}
\end{equation} 
Then the  reduced mean square displacement may be expressed as the time integral of the reduced velocity correlation function (see Eq. \eqref{drviaKvgen}):
\begin{equation}
\left< \left( \Delta \hat{r}(\tau) \right)^2  \right>  = 2 \int_0^{\tau} d\tau_1 \int_{\tau_1}^{\tau} d\tau_2 \, \, \frac32 
\exp \left( - \frac{\tau_2 -\tau_1}{\hat{\tau}_v} \right) \, .    
\label{eq:reduc_r2}
\end{equation} 
Consider now the time derivative of the reduced mean square displacement at time $\tau \gg \hat{\tau}_v$,
\begin{equation}
  \begin{split}
    \frac{d}{d \tau} \left<  \Delta \hat{r}^2   \right>  &= 3 \int_0^{\tau} d\tau_1 
    \exp \left( - \frac{\tau -\tau_1}{\hat{\tau}_v} \right)\\
& =3 \hat{\tau}_v \left[ 1- \exp \left(-\frac{\tau}{\hat{\tau}_v} \right) \right] \simeq 3 \hat{\tau}_v \, , 
\end{split}
\label{eq:dr2scal_dtau}
\end{equation} 
which reads in unscaled variables 
\begin{equation}
  \begin{split}
    \frac{d}{d t} & \left<  \Delta r^2 (t)  \right>  
    = \frac{r_0^2}{\tau_c} \frac{d}{d \tau} \left<  \Delta \hat{r}^2 (\tau)  \right> 
    = \frac{r_0^2}{\tau_c^2} \, 3 \tau_v  \\[0.2cm]
    & = \frac{v_T^2(t) \tau_c^2 (t)}{\tau_c^2(t)} \, \, 3 \tau_v 
    = 6 \frac{T(t)}{m} \tau_v 
    = 6 D(t) \, ,
  \end{split}
  \label{eq:dr2_dtau}
\end{equation} 
where we take into account that $\hat{\tau}_v = \tau_v/\tau_c$ and $v_T^2=2T/m$. Hence, $D= \tau_v T/m$ and, therefore,
\begin{equation}
  D(t)  = \frac{T}{m} \left( \tau_{v, \,{\rm ad}}^{-1}(t) - \frac12 \zeta (t) \right)^{-1} \, .
\label{eq:D_byond_ad}
\end{equation}
To find $ \tau_{v, \,{\rm ad}}$ we evaluate $\left< \vec{v}_1 \cdot  \hat{T}_{12} \vec{v}_1 \right>_t$ using the relation  
\begin{equation}
\label{v12Tv}
\left< \vec{v}_1 \cdot \hat{T}_{12} \, \vec{v}_1 \right>     
     = \frac12 \left< \vec{v}_{12} \cdot \hat{T}_{12} \, \vec{v}_1 \right>  \, ,  
\end{equation} 
which follows from the symmetry properties of the binary collision operator, Eq. \eqref{eq:Tij}. We also use 
\begin{equation}
  \label{eq:v12bv1}
  \begin{split}
    \vec{v}_{12} \cdot \left( \hat{b}_{ij}^{\vec{e}} -1 \right) \vec{v}_1 
    & = \vec{v}_{12} \cdot \left( \vec{v}^{\,\prime}_1-\vec{v}_1 \right) \\
    & = - \frac12 \left( 1 + \varepsilon \right) \left( \vec{v}_{12} \cdot \vec{e}\, \right)^2\,,
  \end{split}
\end{equation}
which in its turn follows from the definition of the operator $\hat{b}_{ij}^{\vec{e}}$, Eq. \eqref{eq:defofbe}. Using  Eq. \eqref{v12Tv} and substituting Eq. \eqref{eq:v12bv1} into the definition of $\tau_{v,\,{\rm ad}}$ in Eq. \eqref{eq:vTv_zeta}, we obtain the velocity relaxation time in adiabatic approximation:
\begin{widetext}
\begin{multline}
\label{eq:tauvgen}
       \tau_{v,\,{\rm ad}}^{-1}(t) 
       = - \left[ \left< v_1^2 \right>_t \right]^{-1} \left( N -1 \right) \frac12   \left< \vec{v}_{12} \cdot \hat{T}_{12} \, \vec{v}_1 \right>_t \\
       =   \frac14 \left[ \frac{3T(t)}{m} \right]^{-1} \left( N -1 \right) \int d\vec{r}_1 \ldots d\vec{r}_N 
            \int d\vec{v}_1 \ldots d\vec{v}_N \rho \left( t \right) 
         \sigma^2  \int d\vec{e} \, \Theta\left(-\vec{v}_{12} \cdot \vec{e} \, \right) \left| \vec{v}_{12} \cdot \vec{e} \, \right| \delta\left(\vec{r}_{12}-\sigma \vec{e} \, \right) 
             \left( 1 + \varepsilon \right) \left( \vec{v}_{12} \cdot \vec{e} \, \right)^2 \, .
\end{multline}
This equation may be evaluated in the same way as the cooling coefficient $\zeta$ in Eqs. \eqtoeqref{TviaTij1}{eq:dTdtcanon}. The result reads
\begin{equation}
  \label{eq:tauvgen2}
  \tau_{v,\,{\rm ad}}^{-1}(t) 
  =   \frac16 v_T(t) g_2(\sigma) \sigma^2 n \int d\vec{c}_1 d\vec{c}_2 \int d\vec{e}  \, \Theta\left(-\vec{c}_{12} \cdot \vec{e} \, \right) \left| \vec{c}_{12} \cdot \vec{e} \, \right| 
  \tilde{f}\left(\vec{c}_1, t\right)\tilde{f}\left(\vec{c}_2, t\right) \left( 1 + \varepsilon \right) \left( \vec{c}_{12} \cdot \vec{e} \, \right)^2 \, ,  
\end{equation}
\end{widetext}
which again is a kinetic integral of the form given in Eq. \eqref{eq:AppType}. Using Eq. \eqref{eq:VDF_HCS} for the distribution function $\tilde{f}\left(c, t\right)$, we can evaluate this kinetic integral by the scheme explained in \cite{PoeschelBrilliantovLNP:2003,BrilliantovPoeschelOUP} and find  
\begin{equation}
\label{eq:tauvad}
\tau_{v,\,{\rm ad}}^{-1}(t) = \frac{\left(1 +\varepsilon \right) }{3} \left( 1+ \frac{3a_2}{32} \right)^2 \tau_c^{-1}(t) \, , 
\end{equation} 
which together with Eqs. \eqref{eq:dTdtcanon} and \eqref{eq:mu2} for the cooling coefficient $\zeta$ leads to the final expression for the self-diffusion coefficient:
\begin{equation}
D(t)  = \frac{4 D_0(t)}{\left( 1 + \varepsilon \right)^2 \left( 1 + \frac{3}{32}a_2 \right)^2} \, , 
\label{eq:D_b_ad_final}
\end{equation} 
where $D_0(t)$ is the Enskog self-diffusion coefficient, 
\begin{equation}
\label{DEnskog}
D_0^{-1}(t)= \frac83 \sigma^2 g_2(\sigma) n  \sqrt{ \frac{\pi m}{T(t) }} \, .
\end{equation}

\section{Chapman-Enskog approach for the diffusion coefficient for $\mathbf{\varepsilon}$\lowercase{=const.}}

We start from the Boltzmann equation for the distribution function of tracers $f_s(\vec{v}, \vec{r},t)$ \cite{BreyRuizMonteroCuberoGarcia:2000,BrilliantovPoeschelOUP}
\begin{equation}
\label{eq:collin_trac}
\begin{split}
&\left(  \frac{\partial}{\partial t} +  \vec{v} \cdot \vec{\nabla} \right) f_s\left(\vec{v}_1,t\right)\\ 
&~~~=g_2\left(\sigma\right)\sigma^2 \int d \vec{v}_2 \int d\vec{e} \, 
\Theta\left(-\vec{v}_{12} \cdot \vec{e} \, \right) \left|\vec{v}_{12} \cdot \vec{e} \,\right| \times\\
&~~~~~~\times \left\{\frac{1}{\varepsilon^2} f_s\left(\vec{v}^{\,\prime\prime}_1,t\right)f\left(\vec{v}^{\,\prime\prime}_2,t\right)-f_s\left(\vec{v}_1,t\right)f\left(\vec{v}_2,t\right) \right\} \\ 
&~~~\equiv g_2\left(\sigma\right)I(f,f_s) \, ,  
\end{split}
\end{equation}
where $I(f,f_s)$ abbreviates the collision integral, $f(\vec{v},t) $ is the velocity distribution function of the gas particles, and the index {\em s} refers to the tracer particles.  In Eq. \eqref{eq:collin_trac}, $\vec{v}^{\,\prime\prime}_1$ and $\vec{v}^{\,\prime\prime}_2$ denote the velocities of particles after an inverse collision, 
\begin{equation}
  \label{eq:inversecoll}
  \begin{split}
    \vec{v}^{\,\prime\prime}_{1}&=\vec{v}_1-\frac{1+\varepsilon}{2 \varepsilon} \left(\vec{v}_{12} \cdot \vec{e} \, \right) \vec{e}\\
    \vec{v}^{\,\prime\prime}_{2}&=\vec{v}_2+\frac{1+\varepsilon}{2 \varepsilon} \left(\vec{v}_{12} \cdot \vec{e} \, \right) \vec{e} \,. 
  \end{split}
\end{equation}
The corresponding direct collision according to Eq. \eqref{eq:bincoll}, thus, turns the initial velocities $(\vec{v}^{\,\prime\prime}_1,\vec{v}^{\,\prime\prime}_2)$ into $(\vec{v}_1, \vec{v}_2)$.

The factors in the integrand at the right-hand side of Eq. \eqref{eq:collin_trac} have their conventional meaning: $\left|\vec{v}_{12} \cdot \vec{e} \,\right|$ is the length of the collision cylinder, $\Theta\left(-\vec{v}_{12} \cdot \vec{e} \, \right)$ selects only approaching particles (since particles departing from each other do not collide) and $g_2\left(\sigma\right)$ accounts for the increased collision frequency due to the finite diameter of the particles and the corresponding excluded volume. Since the concentration of tracers is small, one can assume that they do not affect the velocity distribution function of the gas particles, which is given by the solution, Eq. \eqref{eq:VDF_HCS}, for the homogeneous cooling state. 

Solving the Boltzmann equation for $f_s \left( \vec{r},\vec{v},t \right)$, which with a given function  $f \left(\vec{v},t \right)$ is also called Boltzmann-Lorentz equation, one finds the diffusion flux 
\begin{equation}
\label{eq:J_s_def}
\vec{J}_s \left( \vec{r}, t \right) = \int d\vec{v} \, \vec{v} f_s \left( \vec{r},\vec{v},t \right)  \, ,   
\end{equation}
and then, knowing the concentration gradient, the  diffusion coefficient from the macroscopic equation \eqref{eq:Js_nablans}. 

Solving the linear Boltzmann-Lorentz equation \eqref{eq:collin_trac} is by far not trivial, therefore, the Chapman-Enskog approach has been developed to solve this equation approximatively. This approach is based on two simplifying assumptions: (i) $f_s \left( \vec{r},\vec{v},t \right)$ depends on space and time only trough the macroscopic fields and (ii) $f_s \left( \vec{r},\vec{v},t \right)$ can be expanded in terms of the field gradients. In the gradient expansion 
\begin{equation}
\label{eq:ChEns_fs}
f_s(\vec{r},\vec{v},t)= f_s^{(0)}(\vec{r},\vec{v},t) + \lambda f_s^{(1)}(\vec{r},\vec{v},t) + \lambda^2 f_s^{(2)}(\vec{r},\vec{v},t)\ldots \, 
\end{equation}
a formal parameter $\lambda$ is introduced, which indicates the power of the field gradient. At the end of the computation we set $\lambda=1$.

The first term at the right-hand side, $f_s^{(0)}(\vec{r},\vec{v},t)$, is the distribution function of the uniform system. Since the tagged particles are mechanically identical to the rest of the particles,  $f_s^{(0)}(\vec{r},\vec{v},t)$ is just proportional to  the distribution function of the embedding gas:   
\begin{equation}
\label{eq:fs_homog}
f^{(0)}_s \left( \vec{r}, t \right) = \frac{n_s \left(\vec{r}, t \right) }{n} f(\vec{v},t) \,.  
\end{equation}
Substituting Eq. \eqref{eq:ChEns_fs} into the Boltzmann-Lorentz equation \eqref{eq:collin_trac} and collecting terms of the same order of $\lambda$ (that is of the same order in the gradients) we obtain successive equations for $f_s^{(0)}(\vec{r},\vec{v},t)$, $ f_s^{(1)}(\vec{r},\vec{v},t)$, etc. The zeroth-order equation in the gradients yields $f_s^{(0)}(\vec{r},\vec{v},t)$ for the homogeneous cooling state in accordance with Eq. \eqref{eq:fs_homog}. The first-order equation reads
\begin{equation}
\label{eq:Boltz1_tagg}
\frac{\partial^{(0)} f_s^{(1)}}{\partial t} +\frac{\partial^{(1)} f_s^{(0)}}{\partial t} + \vec{v} \cdot \vec{\nabla} f_s^{(0)} = g_2(\sigma) I \left( f, f_s^{(1)} \right) \,, 
\end{equation}
where $ \partial^{(k)} f / \partial t $ indicates that only terms of $k$-th order with respect to the field gradients are taken into account when the time derivative of some function $f$ is computed. For the case of interest the macroscopic fields are $n_s \left( \vec{r},t \right)$ and $T\left( \vec{r},t \right)$, which satisfy in the homogeneous cooling state the equations 
\begin{equation}
\label{eq:dndt_dTdt_lamb}
\begin{split}
    \frac{\partial n_s}{\partial t} &  = \left( \frac{\partial^{(0)}}{\partial t} + \lambda \frac{\partial^{(1)}}{\partial t} + \lambda^2 \frac{\partial^{(2)}}{\partial t} + \ldots \right) n_s 
                                       =\lambda^2 D \nabla^2 n_s    \\
    \frac{\partial T  }{\partial t} &  = \left( \frac{\partial^{(0)}}{\partial t} + \lambda \frac{\partial^{(1)}}{\partial t} + \lambda^2 \frac{\partial^{(2)}}{\partial t} + \ldots \right) T  
                                       = - \zeta T \, .
\end{split}
\end{equation} 
Comparing the terms in $\lambda^0$ we find
\begin{equation}
\frac{\partial^{(0)} n_s}{\partial t} =0\,,\quad\quad \frac{\partial^{(0)} T}{\partial t} = -\zeta T \, ,  
\end{equation}
which yields for the first term in the left-hand side of Eq. \eqref{eq:Boltz1_tagg}
\begin{equation}
  \label{eq:dt0f_s1}
    \frac{\partial^{(0)} f_s^{(1)}}{\partial t} = \frac{\partial^{(0)} n_s}{\partial t} \, \frac{\partial f_s^{(1)} }{\partial n_s} + \frac{\partial^{(0)} T}{\partial t}  \, \frac{\partial f_s^{(1)} }{\partial T } = -\zeta T \frac{\partial f_s^{(1)}}{\partial T} \,.  
\end{equation}
Similarly, taking into account 
\begin{equation}
  \frac{\partial^{(1)} n_s}{\partial t}=0\,,\quad\quad \frac{\partial^{(1)} T}{\partial t}=0\,,    
\end{equation}
according to  Eq. \eqref{eq:dndt_dTdt_lamb},  we obtain the second term in the left-hand side of Eq. \eqref{eq:Boltz1_tagg},
\begin{equation}
\label{eq:dt1f_s0}
\frac{\partial^{(1)} f_s^{(0)}}{\partial t} = \frac{\partial^{(1)} n_s}{\partial t} \, \frac{\partial f_s^{(0)} }{\partial n_s} 
                            + \frac{\partial^{(1)} T}{\partial t}  \, \frac{\partial f_s^{(0)} }{\partial T }
                           = 0 \, .  
\end{equation}
Using Eq. \eqref{eq:fs_homog} with constant $n$ and $f$, we write the last term in the left-hand side of Eq. \eqref{eq:Boltz1_tagg},
\begin{equation}
\label{eq:nablafs0}
\vec{v} \cdot \vec{\nabla} f_s^{(0)} = \frac{f_s^{(0)} }{n_s} \, \vec{v} \cdot \vec{\nabla} n_s  
                           = \frac{1}{n} \left( \vec{v} \cdot \vec{\nabla} n_s \right) f\,,
\end{equation}
and obtain finally the equation for $f_s^{(1)}$:
\begin{equation}
\label{eq:forf1s}
\zeta T \frac{\partial f_s^{(1)}}{\partial T} + g_2(\sigma) I \left( f, f_s^{(1)} \right) = \frac{1}{n} \left( \vec{v} \cdot \vec{\nabla} n_s \right) f   \, .  
\end{equation}
We search for the  solution of Eq. \eqref{eq:forf1s} in the form
\begin{equation}
\label{eq:form_of_f1s}
 f_s^{(1)} = \vec{G} \left(\vec{v} \, \right) \cdot  \vec{\nabla} n_s \left( \vec{r}, t \right) \,, 
\end{equation}
which implies with Eq. \eqref{eq:J_s_def} the diffusion flux
\begin{equation}
\label{eq:D_via_Gv}
\vec{J}_s=\int  \vec{v} \left( \vec{G} \left(\vec{v} \, \right) \cdot  \vec{\nabla} n_s \right) d \vec{v}
       =-D  \vec{\nabla} n_s \, ,   
\end{equation}
where we take into account 
\begin{equation}
\int \vec{v} f_s^{(0)} d \vec{v} =0
\end{equation}
due to the isotropy of $f_s^{(0)}(|\vec{v}\,|)$. From Eq. \eqref{eq:D_via_Gv} follows
\begin{equation}
\int v_i G_j d \vec{v} =-\delta_{ij} D  
\end{equation}
and summing over $i=j$ yields
\begin{equation}
\label{eq:D_viaGvfinal}
D= - \frac13 \int d \vec{v} \, \vec{v} \cdot \vec{G} \,.  
\end{equation} 
We substitute $f_s^{(1)}$ from Eq. \eqref{eq:form_of_f1s} into  Eq. \eqref{eq:forf1s} discarding the factor $\vec{\nabla} n_s$. Then we multiply it by $\frac13 \vec{v}$ and integrate over $\vec{v}$ to obtain
\begin{multline}
\label{eq:Gv_int_overv}
\zeta T \frac{\partial}{\partial T} \frac13 \int d \vec{v} \, \vec{v} \cdot \vec{G} \left(\vec{v} \, \right) 
+ \frac{g_2(\sigma)}{3} \int d \vec{v} \, \vec{v} \cdot I \left( f, \vec{G} \right) \\
= \frac{2}{3nm} \int d\vec{v} \, \frac{mv^2}{2} f(v) \, .  
\end{multline}
The first term in the left-hand side equals $ - \zeta T \partial D / \partial T$, while the right-hand side equals $T/m$, according to Eq. \eqref{eq:def_T}. The structure of Eq. \eqref{eq:forf1s} suggests the Ansatz
\begin{equation}
\label{eq:G_via_vfb0}
\vec{G} \left(\vec{v} \, \right) \propto \vec{v} f(v) = b \vec{v} f(v)  \, ,  
\end{equation}
where the constant $b$ is to be determined from the above equation. With this Ansatz the second term in the left-hand side of Eq. \eqref{eq:Gv_int_overv} may be written as  
\begin{widetext}
  \begin{equation}
    \label{eq:int_vI_Gf1} 
    \begin{split}
      \frac{g_2(\sigma)}{3} & \int d\vec{v}_1  \vec{v}_1  I\left(f, \vec{G} \right) 
      = \frac{g_2(\sigma)}{3} \frac{b}{2} \int d\vec{v}_1 d\vec{v}_2  \int d\vec{e} \, \Theta \left(-\vec{v}_{12} \cdot \vec{e} \, \right) \left|\vec{v}_{12} \cdot \vec{e} \, \right| f\left(\vec{v}_1 \right)f\left(\vec{v}_2 \right) \times \left( \vec{v}_1 - \vec{v}_2 \right) \cdot \left( \vec{v}^{\,\prime}_1 - \vec{v}_1 \right) \\
      &      =-\frac{bT}{6m}n^2\sigma^2 g_2(\sigma) v_T \int d\vec{c}_1 d\vec{c}_2 \int d\vec{e}  \, \Theta \left(-\vec{c}_{12} \cdot \vec{e} \, \right)
      \left|\vec{c}_{12} \cdot \vec{e} \, \right| \tilde{f}\left(\vec{c}_1 \right) \tilde{f}\left(\vec{c}_2 \right)\left( 1 + \varepsilon \right) \left( \vec{c}_{12} \cdot \vec{e} \, \right)^2 = -\frac{bTn}{m}  \tau_{v,\,{\rm ad}}^{-1} \, . 
    \end{split}
    \end{equation}
To evaluate the last expression we use the main property of the collision integral \cite{BrilliantovPoeschelOUP}, 
\begin{equation}
\label{eq:CollIntMainPro}
\int d\vec{v}_1 \psi \left(\vec{v}_1\right) I\left(f,f\right) = \frac{\sigma^2}{2} \int d\vec{v}_1d\vec{v}_2 \int d\vec{e}\,
\Theta\left(-\vec{v}_{12} \cdot \vec{e}\,\right) \left|\vec{v}_{12} \cdot \vec{e}\,\right|
 f\left(\vec{v}_1, t\right)f\left(\vec{v}_2, t\right) \Delta \left[ \psi\left( \vec{v}_1\right)+
\psi\left( \vec{v}_2\right) \right] \, ,
\end{equation}
\end{widetext}
the relation 
\begin{equation}
\label{eq:v2av2_v1av1_eps}
\vec{v}^{\,\prime}_2 -\vec{v}_2 = - \left( \vec{v}^{\,\prime}_1 - \vec{v}_1 \right) = \frac{1+\varepsilon}{2} \left( \vec{v}_{12} \cdot \vec{e}\, \right) \vec{e}  \, ,   
\end{equation}
which follows from the collision law Eq. \eqref{eq:bincoll} and the symmetry of the expression with respect to the exchange of the particles indices $1 \leftrightarrow 2$. In Eq. \eqref{eq:int_vI_Gf1} we also use the scaled distribution functions $\tilde{f}\left(\vec{c}_1\right)$ and $\tilde{f}\left(\vec{c}_2\right)$ and the definition of the adiabatic relaxation time $\tau_{v,\,{\rm ad}}$, Eq. \eqref{eq:tauvgen2}. Taking finally into account that 
\begin{equation}
\label{eq:D_viab0}
   D = -\frac{b}{3} \int d \vec{v} \, \vec{v} \cdot \vec{v} f(v) 
     = -  \frac{bTn}{m}\, ,   
\end{equation}
according to Eqs. \eqeqref{eq:D_viaGvfinal}{eq:G_via_vfb0} and \eqref{eq:def_T}, we recast Eq. \eqref{eq:Gv_int_overv} into the form
\begin{equation}
\label{eq:forD_final}
-\zeta T \frac{\partial D}{\partial T} +  D \tau_{v,\,{\rm ad}}^{-1} = \frac{T}{m}  \,.  
\end{equation}
Simple arguments show that the diffusion coefficient scales as $D \propto \sqrt{T}$: According to  Eqs. \eqref{eq:tauvgen2} and \eqref{eq:dTdtcanon} the quantities $\tau_{v,\,{\rm ad}}$ and  $\zeta T$ scale as 
\begin{equation}
  \begin{split}
    \tau_{v,\,{\rm ad}}^{-1} &\propto v_T \propto \sqrt{T}\\
    \zeta T \frac{\partial}{\partial T}  &\propto \zeta \propto v_T \propto \sqrt{T}\,.
  \end{split}
\end{equation}
At the same time, the right-hand side of Eq. \eqref{eq:forD_final} scales as $\propto T$, which implies that  $D \propto \sqrt{T}$. Therefore 
\begin{equation}
T \frac{\partial D}{\partial T} = \frac{D}{2}  
\end{equation}
and we obtain the solution of Eq. \eqref{eq:forD_final}, 
\begin{equation}
D(t)  = \frac{T}{m} \left[ \tau_{v, \,{\rm ad}}^{-1}(t) - \frac12 \zeta (t) \right]^{-1} \, , 
\label{eq:D_byond_ad1}
\end{equation} 
which coincides with Eq. \eqref{eq:D_byond_ad}. Substituting $\tau_{v,\,{\rm ad}}$ and $\zeta$ given by Eqs. \eqref{eq:tauvad} and \eqref{eq:dTdtcanon} into Eq. \eqref{eq:D_byond_ad1} we obtain again the diffusion coefficient, Eq. \eqref{eq:D_b_ad_final}. Hence the Chapman-Enskog approach yields for the case $\varepsilon=$const. the same result for the diffusion coefficient as the Green-Kubo approach. 

\section{Diffusion coefficient for impact-velocity dependent coefficient of restitution}

\subsection{Dynamics of a granular gas with impact-velocity dependent coefficient of restitution}
\label{sec:epsdiss}

In the previous sections it was assumed that the coefficient of restitution, which determines the collision dynamics in granular gases, is constant, i.e., independent of the impact velocity. Now we consider collision models, where this dependence is taken into account. 

For both models addressed above, the stepwise model due to Eq. \eqref{eq:epsstep} and the realistic dependence for viscoelastic spheres, Eq. \eqref{eq:epsvisco}, there exist a characteristic velocity, namely $g^*$ for the former case and $\gamma^{-5}$ for the latter. The existence of this additional characteristic velocity, which is not related to the thermal velocity, leads to serious consequences for the granular gas dynamics: Rescaling time according to the Eq. \eqref{eq:tau_t_viatauc} does not lead to an effective steady-state of the system. Indeed, although the average kinetic energy of particles and their collision frequency is kept constant under this transformation, the value of the characteristic velocity ($g^*$ or $\gamma^{-5}$) in the new units increases with time. This follows from the fact that in the homogeneous cooling state the collision frequency steadily decreases in the laboratory time. Hence the time, measured in the new units, slows down [see Eq. \eqref{eq:tau_t_viatauc}] and the characteristic velocity (fixed in the laboratory time) respectively  grows. Therefore, a steady-state may not be achieved under this transformation, since the characteristic velocity, which determines the collision dynamics, is not a constant in the new  time units. 

Consequently, in contrast to the case $\varepsilon=$const, it is not possible to map the dynamics of a granular gas with impact-velocity dependent coefficient of restitution onto a steady state dynamics of a system whose time slows down. As the result we cannot use Doob's theorem \cite{resibua} and deduce the exponential correlation function under the assumption of a Markovian collision process. This makes application of the Green-Kubo approach much more complicated, unless the adiabatic approximation is adopted   \cite{BrilliantovPoeschelOUP}. The adiabatic approximation assumes that the relaxation rate of temperature, proportional to $\zeta(t)$, Eq. \eqref{eq:dTdtcanon}, is negligibly small as compared to the relaxation rate of the velocity distribution function, $\tau_v(t)^{-1}$. Under this approximation one can still assume 
 exponential form of the velocity correlation function,  since  on the  time scale of the velocity correlation, the system is almost in a steady state \cite{BrilliantovPoeschel:1998d}. 

From the above Eqs. \eqref{eq:zetadef}, \eqref{eq:a2Ctn}, \eqref{eq:mu2} for $\zeta$ and Eqs. \eqref{eq:vTv_zeta}, \eqref{eq:tauvad} for $\tau_v$ follows for the case $\varepsilon = {\rm const.}$: $ \zeta(t) \sim (1-\varepsilon^2) \tau_c(t)^{-1}$ and $\tau_v(t)^{-1} \sim \tau_c(t)^{-1}$, which yields the condition of the validity of the adiabatic approximation,
\begin{equation}
\label{eq:ValidAdiab}
(1-\varepsilon^2) \ll 1 \, .
\end{equation}
The same condition may be obtained for the case of impact-velocity dependent coefficient of restitution. If this condition is not justified it is preferable to use the Chapman-Enskog
approach to compute the diffusion coefficient.

\subsection{Diffusion coefficient for the stepwise impact-velocity dependent coefficient of restitution}

Since the shape of the velocity distribution function does not depend on time for this collision model \cite{PoeschelBrilliantovSchwager:2003} the derivation of the diffusion coefficient by means of the Chapman-Enskog method is identical to the case $\varepsilon=$const, presented in the previous chapter. It leads  thus, to the same Eq. \eqref{eq:forD_final} for the diffusion coefficient. The cooling coefficient is given by Eqs. \eqeqref{eq:zetadef}{eq:mu2_fin} for this model, while $\tau_{v, \,{\rm ad}}^{-1}$ may be found from the general relation, Eq. \eqref{eq:tauvgen2}. Using $\tilde{f}\left(\vec{c}_1 \right)=\phi(c_1)$ and $\tilde{f}\left(\vec{c}_2 \right)=\phi(c_2)$ in Eq. \eqref{eq:tauvgen2}  and Eq. \eqref{eq:epsstep} for the coefficient of restitution,   we again express $\tau_{v, \,{\rm ad}}^{-1}$ as a kinetic integral. Its solution yields (see Appendix \ref{app:zeta})
\begin{equation}
  \label{eq:tauvadEpstep}
  \begin{split}
    \tau_{v,\,{\rm ad}}^{-1}&(t) =   \tau_c^{-1}(t) \times \\
    & \left[ \frac23 +\frac{\varepsilon^*-1}{3} \left(1+ \frac{m g^{* \, 2}}{4T} \right) \exp \left( -\frac{m g^{* \, 2}}{4T}\right)  \right] \, , 
  \end{split}
\end{equation} 
with $ \tau_c^{-1}(t)$ given by Eq. \eqref{tauc01}. To find the diffusion coefficient, the first-order non-homogeneous differential equation \eqref{eq:forD_final} must be solved, which may be always written in quadratures. The resulting expression is rather complicated and not of practical use. In both limits  $T \gg mg^{* \, 2}/4$ and $T \ll  mg^{* \, 2}/4$ the diffusion coefficient scales in the same way as for $\varepsilon=$const, that is, $D \propto \sqrt{T}$. Therefore in these limits Eq. \eqref{eq:D_byond_ad1} stays valid. We extrapolate this dependence for the full interval of temperature, that is, we use the approximation
\begin{equation}
D(t)  \approx \frac{T}{m} \left[ \tau_{v, \,{\rm ad}}^{-1}(t) - \frac12 \zeta (t) \right]^{-1} \, , 
\label{eq:D_byond_approx}
\end{equation} 
with  $\zeta$ and $ \tau_{v, \,{\rm ad}}$  given respectively by Eqs. \eqeqref{eq:zetadef}{eq:mu2_fin} and \eqref{eq:tauvadEpstep}. 

\begin{figure}[htb]
  \centerline{\includegraphics[width=8.5cm,clip]{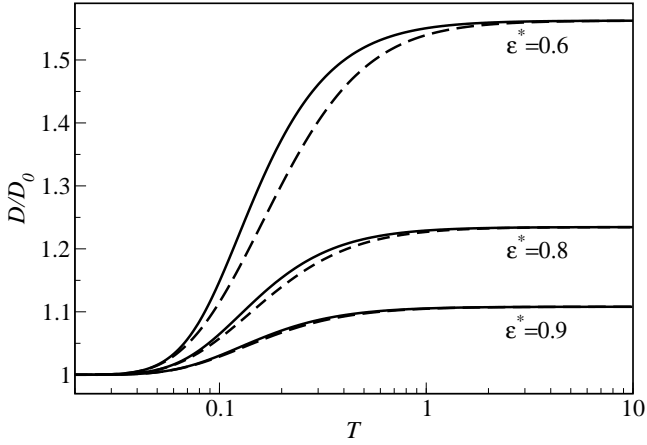}}
  \caption{Reduced self-diffusion coefficient $D/D_0$, where $D_0$ is the Enskog self-diffusion coefficient given by Eq. \eqref{DEnskog}, as a function of temperature for the stepwise impact-velocity dependent coefficient of restitution, Eq. \eqref{eq:epsstep}. Full line: numerical solution of Eq. \eqref{eq:forD_final}; dashed line: approximation given by Eq. \eqref{eq:D_byond_ad1}. The parameters are: $m=1$, $\sigma=1$, $g^*=1$.}
  \label{fig:DiffTEpstep}
\end{figure}

The interesting feature of this model is the relatively sharp dependence of the diffusion
coefficient on the granular temperature: For $T\approx mg^{* \, 2}/4$ the diffusion
coefficients varies exponentially fast with temperature. If $\varepsilon^* = 0.6$ it changes
by about 50$\%$ in a narrow interval of temperature, Fig. \ref{fig:DiffTEpstep}. Note that the
approximation \eqref{eq:D_byond_approx} agrees fairly well with the result obtained by the
numerical solution of  Eq. \eqref{eq:forD_final}, Fig. \ref{fig:DiffTEpstep}. \vspace{0.2cm}

\subsection{Diffusion coefficient for gases of viscoelastic particles}
The application of the Chapman-Enskog method for granular gases of viscoelastic particles is similar to the case $\varepsilon=$const. For gases of viscoelastic particles, however, some complications arise which are related to the additional dependence of the velocity distribution function on time: Here the shape of the velocity distribution function is not invariant, but depends on time via the time-dependent coefficients of the Sonine polynomial expansion \cite{BrilliantovPoeschel:2000visc,BrilliantovPoeschelOUP}. If we keep only two first non-vanishing terms of the Sonine polynomial expansion for the distribution function, we need to account for the time dependence of the second Sonine coefficient $a_2(t)$. Then Eq. \eqref{eq:dt0f_s1} adopts the more general form
\begin{multline}
\label{eq:dt0f_s1_vis}
\frac{\partial^{(0)} f_s^{(1)}}{\partial t} = \frac{\partial^{(0)} n_s}{\partial t} \, \frac{\partial f_s^{(1)} }{\partial n_s} 
                      + \frac{\partial^{(0)} T}{\partial t}  \, \frac{\partial f_s^{(1)} }{\partial T } 
                      + \frac{\partial^{(0)} a_2}{\partial t} \, \frac{\partial f_s^{(1)} }{\partial a_2} \\[0.2cm]
                      = -\zeta T \frac{\partial f_s^{(1)}}{\partial T} + \dot{a}_2 \, \frac{\partial f_s^{(1)} }{\partial a_2} \,,  
\end{multline}
If we  again choose $f_s^{(1)}( \vec{v}\, )$ as given by Eq. \eqref{eq:form_of_f1s} and perform the same steps of analysis which lead from Eq. \eqref{eq:Boltz1_tagg} to Eq. \eqref{eq:Gv_int_overv}, we arrive at
\begin{multline}
\label{eq:Gv_int_visc}
\zeta T \frac{\partial}{\partial T} \frac13 \int d \vec{v} \vec{v} \cdot \vec{G} \left(\vec{v} \, \right)\\ 
  - \frac13 \dot{a}_2 \int d \vec{v} \, \vec{v} \cdot \frac{ \partial \vec{G} }{ \partial a_2} 
+ \frac{g_2(\sigma)}{3} \int d \vec{v} \, \vec{v} \cdot I \left( f, \vec{G} \right) \\
= \frac{2}{3nm} \int d\vec{v} \, \frac{mv^2}{2} f(v) \, .  
\end{multline}
We substitute $\vec{G}(\vec{v}\,)$ by means of Eq. \eqref{eq:G_via_vfb0} with 
\begin{equation}
  \label{eq:unscalVDF}
f(v,t)=f_M(v)\left[ 1 +a_2 (t) S_2\left(c^2\right) \right] \,,     
\end{equation}
where $f_M(v) =(n/v_T^3)\phi (c)$ is the unscaled Maxwell distribution. 

All terms in Eq. \eqref{eq:Gv_int_visc}, except for the second term in the left hand side, have been already evaluated. The remaining term reads
\begin{multline}
\label{eq:int_dG_da2}
\frac{\dot{a}_2}{3} \int d \vec{v} \, \vec{v} \cdot \frac{ d\vec{G} }{d a_2} =  
\frac{\dot{a}_2}{3} b \int d \vec{v} \,  \vec{v} \cdot \vec{v} f_M \, S_2\left(c^2\right) \\
        =  \frac13 \dot{a}_2  b_0nv_T^2 \int d \vec{c} \, \phi(c) c^2 S_2\left(c^2\right)  
        = 0  \,,   
\end{multline}
where we use 
\begin{equation}
\int d \vec{c} \phi(c) c^2 S_2\left(c^2\right) =0\,,  
\end{equation}
which follows directly from the definition of the second Sonine polynomial, Eq. \eqref{eq:VDF_HCS}. Hence we arrive again at Eq. \eqref{eq:forD_final} for the diffusion coefficient as for the case $\varepsilon=\text{const}$: 
\begin{equation}
-\zeta T \frac{\partial D}{\partial T} +  D \tau_{v,\,{\rm ad}}^{-1} = \frac{T}{m}  \,. 
\label{eq:D_byond_ad2}
\end{equation} 
Evaluating the kinetic integral, Eq. \eqref{eq:tauvgen2}, for viscoelastic particles yields $\tau_{v,\,{\rm ad}}^{-1}$ \cite{PoeschelBrilliantovLNP:2003,BrilliantovPoeschelOUP}:
\begin{equation}
\label{tauvvisco}
\tau_{v,\,{\rm ad}}^{-1} =\frac23 \tau_c^{-1}(t) 
\left[ 1+\frac{3}{16}a_2 
- \frac{\omega_0}{4 \sqrt{2\pi}}  \delta^{\prime}(t) \right]\,,
\end{equation}
where $\tau_c^{-1}(t) $ is given by Eq. \eqref{tauc01}, while $a_2(t)$ and $\delta^{\, \prime} (t)$ by Eqs. \eqeqref{eq:a2viscoadiab}{eq:defdelprime}. Hence we note,  that both $\tau_{v,\,{\rm ad}}^{-1}$ as well as $\zeta$ with Eqs. \eqeqref{eq:zetadef}{eq:defmu2Visco} may be written as an expansion in terms of the small parameter $\delta^{\prime}$. 

To solve Eq. \eqref{eq:D_byond_ad2} for the diffusion coefficient we write $D(t)$ as an expansion too,
\begin{equation}
D(t)  = D_0 \left( 1 + \delta^{\prime} \tilde{D}_1 + \delta^{\prime \, 2} \tilde{D}_2 + \cdots \right) \,
\label{eq:D_expand_delta}
\end{equation} 
and substitute this expression into Eq. \eqref{eq:D_byond_ad2} together with equivalent expansions for $\zeta(t)$ and $\tau_{v, \,{\rm ad}}^{-1}(t)$. The resulting equation can be solved perturbatively for each order of $\delta^{\prime}$. Here we give the final result of these straightforward calculations \cite{BrilliantovPoeschel:2003,BrilliantovPoeschelOUP}:
\begin{multline}
  \frac{D(t)}{D_0}  = \left[ 1 + \frac{169}{320} \frac{\omega_0}{\sqrt{2 \pi}} \, \delta^{\prime} \right.\\
  \left. + \left( \frac{2867777}{20480000} \frac{\omega_0^2}{\pi} 
      - \frac{3611}{7040} \frac{\omega_1}{\sqrt{2 \pi}}  \right) \delta^{\prime \, 2}  + \cdots \right]\,,
\label{eq:D_expand_final}
\end{multline} 
where  $D_0$ is  the Enskog coefficient of self-diffusion at the initial temperature $T_0$, 
\begin{equation}
\label{eq:DEnskog_T0}
D_0^{-1}= \frac83 \sigma^2 g_2(\sigma) n  \sqrt{ \frac{\pi m}{T_0 }} \, .
\end{equation}

\section{Conclusion}
We studied diffusion of tracer particles (self-diffusion) in a gas of dissipatively colliding particles. Two main theoretical approaches for the calculation of transport coefficients have been considered: the Green-Kubo and Chapman-Enskog approaches. The Green-Kubo approach is based on the computation of the time integral of the time-correlation functions of dynamical variables, while Chapman-Enskog method is based on the Boltzmann equation. The dissipative collisions of the particles are described by the coefficient of restitution. We addressed three different models for this coefficient: (i) the model of a constant coefficient of restitution, (ii) the model of a simplified (stepwise) dependence of the coefficient of restitution on the impact velocity, and (iii) the realistic model for the coefficient of restitution which follows from the contact mechanics of viscoelastic particles. 

For the case (i), $\varepsilon=$const, the dynamics of the dissipative system may be mapped onto the steady-state dynamics of another gas. This allows for the application of the standard methods of statistical physics, elaborated for molecular gases in thermodynamic equilibrium. In particular the velocity time-correlation function may be straightforwardly evaluated and the diffusion coefficient is correspondingly found. The obtained Green-Kubo diffusion coefficient coincides with that derived using the Chapman-Enskog approach. In contrast, for the case of impact-velocity dependent coefficients of restitution, models (ii) and (iii), there is no simple transformation to map the dynamics of the system onto an effective  steady-state dynamics. This turns the application of the Green-Kubo method either restrictive (e.g. by assuming an adiabatic steady-state), or very complicated and inefficient. Therefore, for gases of particles which interact by an impact-velocity dependent coefficient of restitution, we use the Chapman-Enskog approach to compute the diffusion coefficient. 

Although the Green-Kubo approach is more general, in principle, since it does not assume any form of the time correlation function, while the Chapman-Enskog approach exploits the assumption of molecular chaos, the Chapman-Enskog method appears to be preferable for the application to granular gases. While the latter one is applicable to any collision model, the former one becomes either restrictive or very complicated if the coefficient of restitution depends on the impact velocity. 

\appendix
\section{Second Sonine coefficient}
\label{app:sonine}

To find the velocity distribution function in the homogeneous cooling state we write $f(v)$ in  the rescaled form, Eq. \eqref{eq:defVDFscal}, and substitute it into the Boltzmann equation \eqref{eq:collin_trac} (with the term $\vec{v} \cdot \vec{\nabla}$ omitted and with $f_s$ substituted by $f$). Then this equation may be reduced to two equations, Eq. \eqref{eq:dTdtcanon} for temperature and another one for the scaled distribution function $\tilde{f}\left( \vec{c},t\right)$ \cite{BrilliantovPoeschel:2000visc,BrilliantovPoeschelOUP}: 
\begin{equation}
\label{1geneqveldis}
\frac{\mu_2}{3}
\left(3 + c_1 \frac{\partial}{\partial c_1} \right) \tilde{f}\left( \vec{c}, t\right) +
B^{-1} \frac{\partial}{\partial t} \tilde{f}\left(\vec{c}, t\right) = \tilde{I}\left( \tilde{f}, \tilde{f} \right)\,,
\end{equation}
where 
\begin{equation}
\label{eq:defB}
B=B(t) \equiv v_T(t) g_2 (\sigma) \sigma^2 n \, 
\end{equation}
and $\tilde{I}\left( \tilde{f}, \tilde{f} \right)$ is the dimensionless collision integral, that is, the collision integral Eq. \eqref{eq:collin_trac} written in terms of the scaled distribution functions and scaled velocities.  The analysis shows that on the time scale of the diffusion process -- the hydrodynamic time scale -- the term $B^{-1} \frac{\partial}{\partial t} \tilde{f}$  in Eq. \eqref{1geneqveldis} can  be omitted \cite{BrilliantovPoeschel:2003}. 

The solution for the scaled distribution function $\tilde{f}$ may be found as an expansion around the Maxwell distribution in terms of orthogonal Sonine polynomials $S_p(x)$ 
\cite{GoldshteinShapiro1:1995,NoijeErnst:1998,BrilliantovPoeschel:2000visc}, 
\begin{equation}
\label{eq:Sonexp}
\tilde{f}(c) = \phi (c) \left[ 1 + \sum_{p=1}^{\infty} a_p(t) S_p(c^2) \right]  \,,  
\end{equation}
where $S_0(x)=1$, $S_1(x)=3/2-x$ and $S_2(x)$ is given by Eq. \eqref{eq:S2defin}.

Multiplying both sides of Eq. \eqref{1geneqveldis} by $c_1^{\,p}$, integrating over $ \vec{c}_1$, and using the orthogonality relation for the Sonine polynomials, we obtain an infinite set of equations for the moments $\mu_p$ 
\cite{NoijeErnst:1998,BrilliantovPoeschelStability:2000}     
\begin{equation}
\label{eq:mu2mup}
3 \, \mu_p = \mu_2 \, p \, \left< c^p \right>  \,.  
\qquad p=2,4, \ldots \, ,  
\end{equation} 
Here $\left< c^p \right> \equiv \int d\vec{c} \, c^p \tilde{f}(c)$ are the moments of the velocity distribution function, 
\begin{multline}
\label{eq:mupviaDelta}
\mu_p = -\int c^p \tilde{I} d \vec{c} \\
      =  -\frac12 \int d \vec{c}_1  d \vec{c}_2 \int d \vec{e}\, \Theta\left(-\vec{c}_{12} \cdot \vec{e}\,\right) \left|\vec{c}_{12} \cdot \vec{e}\,\right| \\
          \times \tilde{f}(c_1)\tilde{f}(c_2) \Delta(c_1^p+c_2^p) \, , 
\end{multline}
and we use the main property of the collision integral, Eq. \eqref{eq:CollIntMainPro}. Since $\left< c^p \right>$ and $\mu_p$ are expressed in terms of the Sonine coefficients, the set of equations (\ref{eq:mu2mup}) may be used to determine these coefficients, i.e., to find the velocity distribution function. Assuming a cutoff of the series Eq. (\ref{eq:Sonexp}), i.e. assuming that the Sonine coefficients $a_k$ with $k>k_0$ are negligible, the infinite set of equations (\ref{eq:mu2mup}) turns into a closed finite set of equations for the coefficients $a_k$.

The first Sonine coefficient $a_1=0$, according to the definition of temperature \cite{NoijeErnst:1998,BrilliantovPoeschelStability:2000}. Since  $\left< c^2 \right> =3/2$, the first equation in Eq. \eqref{eq:mu2mup} for $p=2$ is an identity. Assume that $a_3$, $a_4, \ldots$ may be neglected. Then using $\left<c^4\right>=(15/4)(1+a_2)$ which follows from the definition of the second Sonine polynomial, Eq. \eqref{eq:S2defin}, and the approximation $\tilde{f}(c,t)=\phi(c)[1+a_2(t) S_2(c^2)]$, we write Eq. \eqref{eq:mu2mup} for $p=4$:
\begin{equation}
\label{eq:mu4mu2eq}
5 \mu_2(1+a_2) -\mu_4=0 \, . 
\end{equation}
The coefficients $\mu_2$ and $\mu_4$ are functions of $a_2$ and have the form of kinetic integrals, Eq. \eqref{eq:AppType} and may be straightforwardly evaluated by the scheme introduced in \cite{PoeschelBrilliantovLNP:2003,BrilliantovPoeschelOUP}. For the case $\varepsilon=\text{const.}$, the coefficient $\mu_2$ is given by Eq. \eqref{eq:mu2}, while $\mu_4$ reads \cite{NoijeErnst:1998,BrilliantovPoeschelStability:2000,BrilliantovPoeschelOUP}
\begin{equation}
\label{mu4Alin}
\mu_4=4 \sqrt{2 \pi}\left( T_1+a_2T_2   \right)
\end{equation}
with
\begin{equation}
\label{T12}
\begin{split}
T_1&=\frac14\left(1-\varepsilon^2\right)\left( \frac92+\varepsilon^2 \right) \\
T_2&=\frac{3}{128}\left(1-\varepsilon^2\right)\left(69+10\, \varepsilon^2\right)+\frac12\left(1+\varepsilon \right)\,.
\end{split}
\end{equation}
Substituting Eqs. \eqref{mu4Alin} and \eqref{eq:mu2} into Eq. \eqref{eq:mu4mu2eq} we obtain an equation for the second Sonine coefficient $a_2$. Its linear solution is given by Eq. \eqref{eq:a2Ctn}, while the 
next-order  solution is presented in Refs. \cite{BrilliantovPoeschelStability:2000,BrilliantovPoeschelOUP}.

The velocity distribution function for a gas of viscoelastic particles may be found just in the same way: We evaluate $\mu_2$ and $\mu_4$ which have the form of kinetic integrals, substitute them into Eq. \eqref{eq:mu4mu2eq} and solve the resulting equation for the second Sonine coefficient $a_2$. The final result is given in Eq. \eqref{eq:a2viscoadiab}. Details of the derivation can be found in \cite{BrilliantovPoeschel:2000visc,BrilliantovPoeschelOUP}. 

The derivation of the Sonine coefficients, outlined above, is based on the hypothesis of molecular chaos which is the main precondition of the Boltzmann equation. In the coarse of the evolution of the granular gas, correlations between the particles which have been neglected in this approach, may become important. In particular, it was found  in MD simulations \cite{Nakanishi:2003} that accumulating correlations lead to deviations of $a_2$, from the theoretical predictions, e.g. \cite{NoijeErnst:1998}.

\section{Kinetic integrals  for stepwise impact-velocity dependent coefficient of restitution}
\label{app:zeta}

The scheme elaborated in \cite{PoeschelBrilliantovLNP:2003,BrilliantovPoeschelOUP} for the computation of kinetic integrals may not be directly applied for the evaluation of the cooling coefficient $\zeta$ and the relaxation time $\tau_{v,\,{\rm ad}}$ in case of the stepwise impact-velocity dependent coefficient of restitution defined in Eq. \eqref{eq:epsstep}. Therefore, here we outline their evaluation.

By means of Molecular Dynamics simulations it has been shown that in a granular gas with coefficient of restitution given by Eq. \eqref{eq:epsstep}, the second Sonine coefficient is negligible, that is, $a_2 \approx 0$, provided $\varepsilon^*\lesssim 1$. \cite{PoeschelBrilliantovSchwager:2003}. Then with  
\begin{equation}
\label{eq:Epsteprel}
(1+\varepsilon)=2+(\varepsilon^*-1) \Theta\left[ -\tilde{g^*}-\left(\vec{c}_{12} \cdot \vec{e}\,\right) \right], 
\end{equation}
where $\tilde{g^*} \equiv g^*/v_T$, we write the factor in  $\tau_{v,\,{\rm ad}}^{-1}(t)$, which has the form of the kinetic integral as 
\begin{widetext}
\begin{multline}
\label{eq:KIfortau}
 2 \int d\vec{c}_1 d\vec{c}_2 \int  d\vec{e}\, \Theta\left(-\vec{c}_{12} \cdot \vec{e}\, \right)  \left|\vec{c}_{12} \cdot \vec{e}\,\right|
\tilde{f}\left(\vec{c}_1\,\right)\tilde{f}\left(\vec{c}_2\,\right) \left(\vec{c}_{12} \cdot \vec{e}\,\right)^2  \\
 + (\varepsilon^*-1) \int d\vec{c}_1 d\vec{c}_2 \int  d\vec{e}\, \Theta\left(-\vec{c}_{12} \cdot \vec{e}\, \right)  \left|\vec{c}_{12} \cdot \vec{e}\,\right| 
\tilde{f}\left(\vec{c}_1\,\right)\tilde{f}\left(\vec{c}_2\,\right)
 \left(\vec{c}_{12} \cdot \vec{e}\, \right)^2 \Theta \left[ -\tilde{g^*}-\left(\vec{c}_{12} \cdot \vec{e}\,\right) \right] \,.
\end{multline}
\end{widetext}
Changing variables, $\left(\vec{c}_1, \vec{c}_2\right) \to \left(\vec{C}, \vec{c}_{12}\right)$, where $\vec{C} \equiv (\vec{c}_1+\vec{c}_2)/2$ and $\vec{c}_{12}\equiv \vec{c}_1-\vec{c}_2$, we can write 
\begin{equation}
\label{eq:PhiPhi}
\tilde{f}(\vec{c}_1\,)\tilde{f}(\vec{c}_2\,) = \phi(c_{12}) \phi(C)
\end{equation}
with 
\begin{equation}
\label{eq:PhiPhi1}
\begin{split}
&\phi(c_{12}) = \left( 2\pi \right)^{-d/2} \exp \left( -\frac12 c_{12}^2 \right)  \\
&\phi(C)= \left(\frac{2}{\pi} \right)^{d/2} \exp \left( - 2 C^2  \right) \, .
\end{split}
\end{equation}
Substituting Eq. \eqref{eq:PhiPhi} with Eq. \eqref{eq:PhiPhi1} into Eq. \eqref{eq:KIfortau} and taking into account 
that $\int \phi(C) d \vec{C} =1$, we obtain for the second term in Eq. \eqref{eq:KIfortau}
\begin{multline}
\label{eq:mu2der}
\left(\varepsilon^{*} -1 \right) 4 \pi \int_{\tilde{g^*}}^{\infty}   c_{12}^2 \, \phi(c_{12})\, d c_{12} \\
\times \int_0^{\theta_0} \sin \theta \, c_{12}^3 \, \cos^3 \theta \, d \theta \int_0^{2 \pi} d \varphi \, , 
\end{multline}
with $\theta_0 \equiv \tilde{g^*}/c_{12}$. In Eq. \eqref{eq:mu2der} we take into account that the integral vanishes if the normal component of the impact velocity $\left|\vec{c}_{12} \cdot \vec{e}\right| = \left|c_{12} \cos \theta \right|$ is smaller than $\tilde{g^*}$. Simple calculations then yield for the second term in Eq. \eqref{eq:KIfortau}:
\begin{equation}
\label{eq:tauvad2}
\sqrt{2 \pi} \left(\varepsilon^{*} -1 \right)   \left(1+ \frac{m g^{*\, 2}}{4T} \right) 
\exp \left( - \frac{mg^{*\, 2}}{4T} \right)  \, . 
\end{equation}
The first term of the kinetic integral in Eq. \eqref{eq:KIfortau} may be obtained from the second term, taking $g^*=0$ and omitting the factor $\left(\varepsilon^{*} -1 \right)$. Summing up these two terms and multiplying the result by  $(1/6) v_T g_2(\sigma) \sigma^2 n$, according to the definition Eq. \eqref {eq:tauvgen2}, we arrive at Eq. \eqref{eq:tauvadEpstep} for  $\tau_{v,\,{\rm ad}}^{-1}(t)$. 

The cooling coefficient $\zeta$ may be evaluated in a similar way. From the definition of the collision model, Eq. \eqref{eq:epsstep}, and the general collision law, Eq. \eqref{eq:bincoll}, follows
\begin{multline}
\label{eq:DelEps}
\Delta \left( c_1^2 + c_2^2 \right) = -\left(1-\varepsilon^2\right) \left(\vec{c}_{12} \cdot \vec{e}\,\right)^2  \\
         = -\left( 1-\varepsilon^{*\, 2} \right) \left(\vec{c}_{12} \cdot \vec{e}\,\right)^2 \Theta \left[ -\tilde{g^*}-\left(\vec{c}_{12} \cdot \vec{e}\, \right) \right] \,.
\end{multline}
Substituting the latter expression into Eq. \eqref{eq:mu2def} for $\mu_2$, we arrive at the same kinetic integral as for $\tau_{v,\,{\rm ad}}^{-1}(t)$, which eventually leads to the final result Eq. \eqref{eq:mu2_fin}.



\begin{thebibliography}{49}
\expandafter\ifx\csname natexlab\endcsname\relax\def\natexlab#1{#1}\fi
\expandafter\ifx\csname bibnamefont\endcsname\relax
  \def\bibnamefont#1{#1}\fi
\expandafter\ifx\csname bibfnamefont\endcsname\relax
  \def\bibfnamefont#1{#1}\fi
\expandafter\ifx\csname citenamefont\endcsname\relax
  \def\citenamefont#1{#1}\fi
\expandafter\ifx\csname url\endcsname\relax
  \def\url#1{\texttt{#1}}\fi
\expandafter\ifx\csname urlprefix\endcsname\relax\def\urlprefix{URL }\fi
\providecommand{\bibinfo}[2]{#2}
\providecommand{\eprint}[2][]{\url{#2}}

\bibitem[{\citenamefont{Goldhirsch and Zanetti}(1993)}]{GoldhirschZanetti:1993}
\bibinfo{author}{\bibfnamefont{I.}~\bibnamefont{Goldhirsch}} \bibnamefont{and}
  \bibinfo{author}{\bibfnamefont{G.}~\bibnamefont{Zanetti}},
  \bibinfo{journal}{Phys. Rev. Lett.} \textbf{\bibinfo{volume}{70}},
  \bibinfo{pages}{1619} (\bibinfo{year}{1993}).

\bibitem[{\citenamefont{McNamara}(1993)}]{McNamara:1993}
\bibinfo{author}{\bibfnamefont{S.}~\bibnamefont{McNamara}},
  \bibinfo{journal}{Physics of Fluids A} \textbf{\bibinfo{volume}{5}},
  \bibinfo{pages}{3056} (\bibinfo{year}{1993}).

\bibitem[{\citenamefont{Brito and Ernst}(1998)}]{BritoErnst:1998}
\bibinfo{author}{\bibfnamefont{R.}~\bibnamefont{Brito}} \bibnamefont{and}
  \bibinfo{author}{\bibfnamefont{M.~H.} \bibnamefont{Ernst}},
  \bibinfo{journal}{Europhys. Lett.} \textbf{\bibinfo{volume}{43}},
  \bibinfo{pages}{497} (\bibinfo{year}{1998}).

\bibitem[{\citenamefont{Brilliantov and
  P{\"o}schel}(2004)}]{BrilliantovPoeschelOUP}
\bibinfo{author}{\bibfnamefont{N.~V.} \bibnamefont{Brilliantov}}
  \bibnamefont{and}
  \bibinfo{author}{\bibfnamefont{T.}~\bibnamefont{P{\"o}schel}},
  \emph{\bibinfo{title}{Kinetic Theory of Granular Gases}}
  (\bibinfo{publisher}{Oxford University Press}, \bibinfo{address}{Oxford},
  \bibinfo{year}{2004}).

\bibitem[{\citenamefont{Garzo and Montanero}(2004)}]{GarzoMontanero:2004}
\bibinfo{author}{\bibfnamefont{V.}~\bibnamefont{Garzo}} \bibnamefont{and}
  \bibinfo{author}{\bibfnamefont{J.~M.} \bibnamefont{Montanero}},
  \bibinfo{journal}{Phys. Rev. E} \textbf{\bibinfo{volume}{69}},
  \bibinfo{pages}{021301} (\bibinfo{year}{2004}).

\bibitem[{\citenamefont{Barrat and Trizac}(2002)}]{BarratTrizac_GM:2002}
\bibinfo{author}{\bibfnamefont{A.}~\bibnamefont{Barrat}} \bibnamefont{and}
  \bibinfo{author}{\bibfnamefont{E.}~\bibnamefont{Trizac}},
  \bibinfo{journal}{Granular Matter} \textbf{\bibinfo{volume}{4}},
  \bibinfo{pages}{57} (\bibinfo{year}{2002}).

\bibitem[{\citenamefont{Marconi and Puglisi}(2002)}]{Marconi1Puglisi:2002}
\bibinfo{author}{\bibfnamefont{U.~M.~B.} \bibnamefont{Marconi}}
  \bibnamefont{and} \bibinfo{author}{\bibfnamefont{A.}~\bibnamefont{Puglisi}},
  \bibinfo{journal}{Phys. Rev. E} \textbf{\bibinfo{volume}{65}},
  \bibinfo{pages}{051305} (\bibinfo{year}{2002}).

\bibitem[{\citenamefont{Ben-Naim and
  Krapivsky}(2002)}]{BenNaimKrapivsky_EPJ:2002}
\bibinfo{author}{\bibfnamefont{E.}~\bibnamefont{Ben-Naim}} \bibnamefont{and}
  \bibinfo{author}{\bibfnamefont{P.~L.} \bibnamefont{Krapivsky}},
  \bibinfo{journal}{Eur. Phys. J. E} \textbf{\bibinfo{volume}{8}},
  \bibinfo{pages}{507} (\bibinfo{year}{2002}).

\bibitem[{\citenamefont{Clelland and Hrenya}(2002)}]{ClellandHrenya:2002}
\bibinfo{author}{\bibfnamefont{R.}~\bibnamefont{Clelland}} \bibnamefont{and}
  \bibinfo{author}{\bibfnamefont{C.~M.} \bibnamefont{Hrenya}},
  \bibinfo{journal}{Phys. Rev. E} \textbf{\bibinfo{volume}{65}},
  \bibinfo{pages}{031301} (\bibinfo{year}{2002}).

\bibitem[{\citenamefont{Dahl et~al.}(2002)\citenamefont{Dahl, , Hrenya, Garzo,
  and Dufty}}]{Dahletal:2002}
\bibinfo{author}{\bibfnamefont{S.~R.} \bibnamefont{Dahl}}, ,
  \bibinfo{author}{\bibfnamefont{C.~M.} \bibnamefont{Hrenya}},
  \bibinfo{author}{\bibfnamefont{V.}~\bibnamefont{Garzo}}, \bibnamefont{and}
  \bibinfo{author}{\bibfnamefont{J.}~\bibnamefont{Dufty}},
  \bibinfo{journal}{Phys. Rev. E} \textbf{\bibinfo{volume}{66}},
  \bibinfo{pages}{041301} (\bibinfo{year}{2002}).

\bibitem[{\citenamefont{Wang et~al.}(2003)\citenamefont{Wang, Jin, and
  Ma}}]{WangJinMa:2003}
\bibinfo{author}{\bibfnamefont{H.}~\bibnamefont{Wang}},
  \bibinfo{author}{\bibfnamefont{G.}~\bibnamefont{Jin}}, \bibnamefont{and}
  \bibinfo{author}{\bibfnamefont{Y.}~\bibnamefont{Ma}}, \bibinfo{journal}{Phys.
  Rev. E} \textbf{\bibinfo{volume}{68}}, \bibinfo{pages}{031301}
  (\bibinfo{year}{2003}).

\bibitem[{\citenamefont{Feitosa and Menon}(2002)}]{FeitosaMenon:2002}
\bibinfo{author}{\bibfnamefont{K.}~\bibnamefont{Feitosa}} \bibnamefont{and}
  \bibinfo{author}{\bibfnamefont{N.}~\bibnamefont{Menon}},
  \bibinfo{journal}{Phys. Rev. Lett.} \textbf{\bibinfo{volume}{88}},
  \bibinfo{pages}{198301} (\bibinfo{year}{2002}).

\bibitem[{\citenamefont{Wildman and Parker}(2002)}]{WildmanParker:2002}
\bibinfo{author}{\bibfnamefont{R.~D.} \bibnamefont{Wildman}} \bibnamefont{and}
  \bibinfo{author}{\bibfnamefont{D.~J.} \bibnamefont{Parker}},
  \bibinfo{journal}{Phys. Rev. Lett.} \textbf{\bibinfo{volume}{88}},
  \bibinfo{pages}{064301} (\bibinfo{year}{2002}).

\bibitem[{\citenamefont{Santos and
  Dufty}(2001{\natexlab{a}})}]{SantosDufty_PRL:2001}
\bibinfo{author}{\bibfnamefont{A.}~\bibnamefont{Santos}} \bibnamefont{and}
  \bibinfo{author}{\bibfnamefont{J.}~\bibnamefont{Dufty}},
  \bibinfo{journal}{Phys. Rev. Lett.} \textbf{\bibinfo{volume}{86}},
  \bibinfo{pages}{4823} (\bibinfo{year}{2001}{\natexlab{a}}).

\bibitem[{\citenamefont{Santos and
  Dufty}(2001{\natexlab{b}})}]{SantosDufty_PRE:2001}
\bibinfo{author}{\bibfnamefont{A.}~\bibnamefont{Santos}} \bibnamefont{and}
  \bibinfo{author}{\bibfnamefont{J.}~\bibnamefont{Dufty}},
  \bibinfo{journal}{Phys. Rev. E} \textbf{\bibinfo{volume}{64}},
  \bibinfo{pages}{051305} (\bibinfo{year}{2001}{\natexlab{b}}).

\bibitem[{\citenamefont{Bridges et~al.}(1984)\citenamefont{Bridges, Hatzes, and
  Lin}}]{BridgesHatzesLin:1984}
\bibinfo{author}{\bibfnamefont{F.~G.} \bibnamefont{Bridges}},
  \bibinfo{author}{\bibfnamefont{A.}~\bibnamefont{Hatzes}}, \bibnamefont{and}
  \bibinfo{author}{\bibfnamefont{D.~N.~C.} \bibnamefont{Lin}},
  \bibinfo{journal}{Nature} \textbf{\bibinfo{volume}{309}},
  \bibinfo{pages}{333} (\bibinfo{year}{1984}).

\bibitem[{\citenamefont{Kuwabara and Kono}(1987)}]{KuwabaraKono:1987}
\bibinfo{author}{\bibfnamefont{G.}~\bibnamefont{Kuwabara}} \bibnamefont{and}
  \bibinfo{author}{\bibfnamefont{K.}~\bibnamefont{Kono}},
  \bibinfo{journal}{Jpn. J. Appl. Phys.} \textbf{\bibinfo{volume}{26}},
  \bibinfo{pages}{1230} (\bibinfo{year}{1987}).

\bibitem[{\citenamefont{McNamara and Falcon}(2003)}]{McNamaraFalcon:2003}
\bibinfo{author}{\bibfnamefont{S.}~\bibnamefont{McNamara}} \bibnamefont{and}
  \bibinfo{author}{\bibfnamefont{E.}~\bibnamefont{Falcon}}, in
  \cite{PoeschelBrilliantov:2003}, pp. \bibinfo{pages}{297--226}.

\bibitem[{\citenamefont{Tanaka et~al.}(1991)\citenamefont{Tanaka, Ishida, and
  Tsuji}}]{Tanaka}
\bibinfo{author}{\bibfnamefont{T.}~\bibnamefont{Tanaka}},
  \bibinfo{author}{\bibfnamefont{T.}~\bibnamefont{Ishida}}, \bibnamefont{and}
  \bibinfo{author}{\bibfnamefont{Y.}~\bibnamefont{Tsuji}},
  \bibinfo{journal}{Trans. Jap. Soc. Mech. Eng.} \textbf{\bibinfo{volume}{57}},
  \bibinfo{pages}{456} (\bibinfo{year}{1991}).

\bibitem[{\citenamefont{Luding et~al.}(1994)\citenamefont{Luding, Cl\'ement,
  Blumen, Rajchenbach, and Duran}}]{LudingClementBlumenRajchenbachDuran:1994}
\bibinfo{author}{\bibfnamefont{S.}~\bibnamefont{Luding}},
  \bibinfo{author}{\bibfnamefont{E.}~\bibnamefont{Cl\'ement}},
  \bibinfo{author}{\bibfnamefont{A.}~\bibnamefont{Blumen}},
  \bibinfo{author}{\bibfnamefont{J.}~\bibnamefont{Rajchenbach}},
  \bibnamefont{and} \bibinfo{author}{\bibfnamefont{J.}~\bibnamefont{Duran}},
  \bibinfo{journal}{Phys. Rev. E} \textbf{\bibinfo{volume}{50}},
  \bibinfo{pages}{4113} (\bibinfo{year}{1994}).

\bibitem[{\citenamefont{Luding et~al.}(1996)\citenamefont{Luding, Cl\'ement,
  Rajchenbach, and Duran}}]{LudingClementRajchenbachDuran:1996}
\bibinfo{author}{\bibfnamefont{S.}~\bibnamefont{Luding}},
  \bibinfo{author}{\bibfnamefont{E.}~\bibnamefont{Cl\'ement}},
  \bibinfo{author}{\bibfnamefont{J.}~\bibnamefont{Rajchenbach}},
  \bibnamefont{and} \bibinfo{author}{\bibfnamefont{J.}~\bibnamefont{Duran}},
  \bibinfo{journal}{Europhys. Lett.} \textbf{\bibinfo{volume}{36}},
  \bibinfo{pages}{247} (\bibinfo{year}{1996}).

\bibitem[{\citenamefont{Brilliantov et~al.}(1996)\citenamefont{Brilliantov,
  Spahn, Hertzsch, and P\"oschel}}]{BrilliantovSpahnHertzschPoeschel:1994}
\bibinfo{author}{\bibfnamefont{N.~V.} \bibnamefont{Brilliantov}},
  \bibinfo{author}{\bibfnamefont{F.}~\bibnamefont{Spahn}},
  \bibinfo{author}{\bibfnamefont{J.-M.} \bibnamefont{Hertzsch}},
  \bibnamefont{and}
  \bibinfo{author}{\bibfnamefont{T.}~\bibnamefont{P\"oschel}},
  \bibinfo{journal}{Phys. Rev. E} \textbf{\bibinfo{volume}{53}},
  \bibinfo{pages}{5382} (\bibinfo{year}{1996}).

\bibitem[{\citenamefont{Schwager and
  P{\"o}schel}(1998)}]{SchwagerPoeschel:1998}
\bibinfo{author}{\bibfnamefont{T.}~\bibnamefont{Schwager}} \bibnamefont{and}
  \bibinfo{author}{\bibfnamefont{T.}~\bibnamefont{P{\"o}schel}},
  \bibinfo{journal}{Phys. Rev. E} \textbf{\bibinfo{volume}{57}},
  \bibinfo{pages}{650} (\bibinfo{year}{1998}).

\bibitem[{\citenamefont{Morgado and Oppenheim}(1997)}]{MorgadoOppenheim:1997}
\bibinfo{author}{\bibfnamefont{W.~A.~M.} \bibnamefont{Morgado}}
  \bibnamefont{and}
  \bibinfo{author}{\bibfnamefont{I.}~\bibnamefont{Oppenheim}},
  \bibinfo{journal}{Phys. Rev. E} \textbf{\bibinfo{volume}{55}},
  \bibinfo{pages}{1940} (\bibinfo{year}{1997}).

\bibitem[{\citenamefont{Ram\'{\i}rez et~al.}(1999)\citenamefont{Ram\'{\i}rez,
  P{\"o}schel, Brilliantov, and Schwager}}]{Ramirez:1999}
\bibinfo{author}{\bibfnamefont{R.}~\bibnamefont{Ram\'{\i}rez}},
  \bibinfo{author}{\bibfnamefont{T.}~\bibnamefont{P{\"o}schel}},
  \bibinfo{author}{\bibfnamefont{N.~V.} \bibnamefont{Brilliantov}},
  \bibnamefont{and} \bibinfo{author}{\bibfnamefont{T.}~\bibnamefont{Schwager}},
  \bibinfo{journal}{Phys. Rev. E} \textbf{\bibinfo{volume}{60}},
  \bibinfo{pages}{4465} (\bibinfo{year}{1999}).

\bibitem[{\citenamefont{P\"oschel et~al.}(2003)\citenamefont{P\"oschel,
  Brilliantov, and Schwager}}]{PoeschelBrilliantovSchwager:2003}
\bibinfo{author}{\bibfnamefont{T.}~\bibnamefont{P\"oschel}},
  \bibinfo{author}{\bibfnamefont{N.~V.} \bibnamefont{Brilliantov}},
  \bibnamefont{and} \bibinfo{author}{\bibfnamefont{T.}~\bibnamefont{Schwager}},
  \bibinfo{journal}{Physica A} \textbf{\bibinfo{volume}{325}},
  \bibinfo{pages}{274} (\bibinfo{year}{2003}).

\bibitem[{\citenamefont{Nie et~al.}(2002)\citenamefont{Nie, Ben-Naim, and
  Chen}}]{NieBenNaimChen2002}
\bibinfo{author}{\bibfnamefont{X.}~\bibnamefont{Nie}},
  \bibinfo{author}{\bibfnamefont{E.}~\bibnamefont{Ben-Naim}}, \bibnamefont{and}
  \bibinfo{author}{\bibfnamefont{S.~Y.} \bibnamefont{Chen}},
  \bibinfo{journal}{Phys. Rev. Lett.} \textbf{\bibinfo{volume}{89}},
  \bibinfo{pages}{204301} (\bibinfo{year}{2002}).

\bibitem[{\citenamefont{Abramowitz and Stegun}(1965)}]{AbramowitzStegun:1965}
\bibinfo{author}{\bibfnamefont{M.}~\bibnamefont{Abramowitz}} \bibnamefont{and}
  \bibinfo{author}{\bibfnamefont{A.}~\bibnamefont{Stegun}},
  \emph{\bibinfo{title}{Handbook of Mathematical Functions.}}
  (\bibinfo{publisher}{Dover Publications}, \bibinfo{address}{New York},
  \bibinfo{year}{1965}).

\bibitem[{\citenamefont{Resibois and de~Leener}(1977)}]{resibua}
\bibinfo{author}{\bibfnamefont{P.}~\bibnamefont{Resibois}} \bibnamefont{and}
  \bibinfo{author}{\bibfnamefont{M.}~\bibnamefont{de~Leener}},
  \emph{\bibinfo{title}{Classical Kinetic Theory of Fluids}}
  (\bibinfo{publisher}{Wiley \& Sons}, \bibinfo{address}{New York},
  \bibinfo{year}{1977}).

\bibitem[{\citenamefont{Goldhirsch and van Noije}(2000)}]{GoldhirschNoije:2000}
\bibinfo{author}{\bibfnamefont{I.}~\bibnamefont{Goldhirsch}} \bibnamefont{and}
  \bibinfo{author}{\bibfnamefont{T.~P.~C.} \bibnamefont{van Noije}},
  \bibinfo{journal}{Phys. Rev. E} \textbf{\bibinfo{volume}{61}},
  \bibinfo{pages}{3241} (\bibinfo{year}{2000}).

\bibitem[{\citenamefont{Brey et~al.}(2003)\citenamefont{Brey, Dufty, and
  Ruiz-Montero}}]{BreyDuftyRuizMontero:2003}
\bibinfo{author}{\bibfnamefont{J.~J.} \bibnamefont{Brey}},
  \bibinfo{author}{\bibfnamefont{J.~W.} \bibnamefont{Dufty}}, \bibnamefont{and}
  \bibinfo{author}{\bibfnamefont{M.~J.} \bibnamefont{Ruiz-Montero}}, in
  \cite{PoeschelBrilliantov:2003}, pp. \bibinfo{pages}{185--220}.

\bibitem[{\citenamefont{Brey et~al.}(2004)\citenamefont{Brey, , and
  Ruiz-Montero}}]{BreyRuizMonteroPRE:2004}
\bibinfo{author}{\bibfnamefont{J.~J.} \bibnamefont{Brey}}, , \bibnamefont{and}
  \bibinfo{author}{\bibfnamefont{M.~J.} \bibnamefont{Ruiz-Montero}},
  \bibinfo{journal}{Phys. Rev. E} \textbf{\bibinfo{volume}{70}},
  \bibinfo{pages}{051301} (\bibinfo{year}{2004}).

\bibitem[{\citenamefont{Esipov and P\"oschel}(1997)}]{EsipovPoeschel:1995}
\bibinfo{author}{\bibfnamefont{S.~E.} \bibnamefont{Esipov}} \bibnamefont{and}
  \bibinfo{author}{\bibfnamefont{T.}~\bibnamefont{P\"oschel}},
  \bibinfo{journal}{J. Stat. Phys.} \textbf{\bibinfo{volume}{86}},
  \bibinfo{pages}{1385} (\bibinfo{year}{1997}).

\bibitem[{\citenamefont{Brilliantov and
  P\"oschel}(2000{\natexlab{a}})}]{BrilliantovPoeschel:1998d}
\bibinfo{author}{\bibfnamefont{N.~V.} \bibnamefont{Brilliantov}}
  \bibnamefont{and}
  \bibinfo{author}{\bibfnamefont{T.}~\bibnamefont{P\"oschel}},
  \bibinfo{journal}{Phys. Rev. E} \textbf{\bibinfo{volume}{61}},
  \bibinfo{pages}{1716} (\bibinfo{year}{2000}{\natexlab{a}}).

\bibitem[{\citenamefont{Dufty et~al.}(2002)\citenamefont{Dufty, Brey, and
  Lutsko}}]{DuftyBreyLutsko:2002}
\bibinfo{author}{\bibfnamefont{J.}~\bibnamefont{Dufty}},
  \bibinfo{author}{\bibfnamefont{J.~J.} \bibnamefont{Brey}}, \bibnamefont{and}
  \bibinfo{author}{\bibfnamefont{J.}~\bibnamefont{Lutsko}},
  \bibinfo{journal}{Phys. Rev. E} \textbf{\bibinfo{volume}{65}},
  \bibinfo{pages}{051303} (\bibinfo{year}{2002}).

\bibitem[{\citenamefont{Ernst et~al.}(1969)\citenamefont{Ernst, Dorfman, Hoegy,
  and van Leeuwen}}]{ErnstEtAl:1969}
\bibinfo{author}{\bibfnamefont{M.~H.} \bibnamefont{Ernst}},
  \bibinfo{author}{\bibfnamefont{J.~R.} \bibnamefont{Dorfman}},
  \bibinfo{author}{\bibfnamefont{W.~R.} \bibnamefont{Hoegy}}, \bibnamefont{and}
  \bibinfo{author}{\bibfnamefont{J.~M.~J.} \bibnamefont{van Leeuwen}},
  \bibinfo{journal}{Physica A} \textbf{\bibinfo{volume}{45}},
  \bibinfo{pages}{127} (\bibinfo{year}{1969}).

\bibitem[{\citenamefont{Ernst and Dorfman}(1972)}]{ErnstDorfman:1972}
\bibinfo{author}{\bibfnamefont{M.~H.} \bibnamefont{Ernst}} \bibnamefont{and}
  \bibinfo{author}{\bibfnamefont{J.~R.} \bibnamefont{Dorfman}},
  \bibinfo{journal}{Physica A} \textbf{\bibinfo{volume}{61}},
  \bibinfo{pages}{157} (\bibinfo{year}{1972}).

\bibitem[{\citenamefont{Carnahan and Starling}(1969)}]{CarnahanStarling}
\bibinfo{author}{\bibfnamefont{N.~F.} \bibnamefont{Carnahan}} \bibnamefont{and}
  \bibinfo{author}{\bibfnamefont{K.~E.} \bibnamefont{Starling}},
  \bibinfo{journal}{J. Chem. Phys.} \textbf{\bibinfo{volume}{51}},
  \bibinfo{pages}{635} (\bibinfo{year}{1969}).

\bibitem[{\citenamefont{Goldshtein and
  Shapiro}(1995)}]{GoldshteinShapiro1:1995}
\bibinfo{author}{\bibfnamefont{A.}~\bibnamefont{Goldshtein}} \bibnamefont{and}
  \bibinfo{author}{\bibfnamefont{M.}~\bibnamefont{Shapiro}},
  \bibinfo{journal}{J. Fluid Mech.} \textbf{\bibinfo{volume}{282}},
  \bibinfo{pages}{75} (\bibinfo{year}{1995}).

\bibitem[{\citenamefont{van Noije and Ernst}(1998)}]{NoijeErnst:1998}
\bibinfo{author}{\bibfnamefont{T.~P.~C.} \bibnamefont{van Noije}}
  \bibnamefont{and} \bibinfo{author}{\bibfnamefont{M.~H.} \bibnamefont{Ernst}},
  \bibinfo{journal}{Granular Matter} \textbf{\bibinfo{volume}{1}},
  \bibinfo{pages}{57} (\bibinfo{year}{1998}).

\bibitem[{\citenamefont{P\"oschel and
  Brilliantov}(2003{\natexlab{a}})}]{PoeschelBrilliantovLNP:2003}
\bibinfo{author}{\bibfnamefont{T.}~\bibnamefont{P\"oschel}} \bibnamefont{and}
  \bibinfo{author}{\bibfnamefont{N.~V.} \bibnamefont{Brilliantov}}, in
  \cite{PoeschelBrilliantov:2003}, pp. \bibinfo{pages}{129--160}.

\bibitem[{\citenamefont{Brilliantov and
  P\"oschel}(2000{\natexlab{b}})}]{BrilliantovPoeschelStability:2000}
\bibinfo{author}{\bibfnamefont{N.~V.} \bibnamefont{Brilliantov}}
  \bibnamefont{and}
  \bibinfo{author}{\bibfnamefont{T.}~\bibnamefont{P\"oschel}},
  \bibinfo{journal}{Phys. Rev. E} \textbf{\bibinfo{volume}{61}},
  \bibinfo{pages}{2809} (\bibinfo{year}{2000}{\natexlab{b}}).

\bibitem[{\citenamefont{Huthmann et~al.}(2000)\citenamefont{Huthmann, Orza, and
  Brito}}]{HuthmannOrzaBrito:2000}
\bibinfo{author}{\bibfnamefont{M.}~\bibnamefont{Huthmann}},
  \bibinfo{author}{\bibfnamefont{J.}~\bibnamefont{Orza}}, \bibnamefont{and}
  \bibinfo{author}{\bibfnamefont{R.}~\bibnamefont{Brito}},
  \bibinfo{journal}{Granular Matter} \textbf{\bibinfo{volume}{2}},
  \bibinfo{pages}{189} (\bibinfo{year}{2000}).

\bibitem[{\citenamefont{Nakanishi}(2003)}]{Nakanishi:2003}
\bibinfo{author}{\bibfnamefont{H.}~\bibnamefont{Nakanishi}},
  \bibinfo{journal}{Phys. Rev. E} \textbf{\bibinfo{volume}{67}},
  \bibinfo{pages}{010301(R)} (\bibinfo{year}{2003}).

\bibitem[{\citenamefont{Brilliantov and
  P\"oschel}(2003)}]{BrilliantovPoeschel:2003}
\bibinfo{author}{\bibfnamefont{N.~V.} \bibnamefont{Brilliantov}}
  \bibnamefont{and}
  \bibinfo{author}{\bibfnamefont{T.}~\bibnamefont{P\"oschel}},
  \bibinfo{journal}{Phys. Rev. E} \textbf{\bibinfo{volume}{67}},
  \bibinfo{pages}{061304} (\bibinfo{year}{2003}).

\bibitem[{\citenamefont{Brilliantov and
  P\"oschel}(2000{\natexlab{c}})}]{BrilliantovPoeschel:2000visc}
\bibinfo{author}{\bibfnamefont{N.~V.} \bibnamefont{Brilliantov}}
  \bibnamefont{and}
  \bibinfo{author}{\bibfnamefont{T.}~\bibnamefont{P\"oschel}},
  \bibinfo{journal}{Phys. Rev. E} \textbf{\bibinfo{volume}{61}},
  \bibinfo{pages}{5573} (\bibinfo{year}{2000}{\natexlab{c}}).

\bibitem[{\citenamefont{Brey et~al.}(1999)\citenamefont{Brey, Ruiz-Montero, and
  Garcia-Rojo}}]{BreyRuizMonteroGarciaRojo:1999}
\bibinfo{author}{\bibfnamefont{J.~J.} \bibnamefont{Brey}},
  \bibinfo{author}{\bibfnamefont{M.~J.} \bibnamefont{Ruiz-Montero}},
  \bibnamefont{and}
  \bibinfo{author}{\bibfnamefont{R.}~\bibnamefont{Garcia-Rojo}},
  \bibinfo{journal}{Phys. Rev. E} \textbf{\bibinfo{volume}{60}},
  \bibinfo{pages}{7174} (\bibinfo{year}{1999}).

\bibitem[{\citenamefont{Brey et~al.}(2000)\citenamefont{Brey, Ruiz-Montero,
  Cubero, and Garcia-Rojo}}]{BreyRuizMonteroCuberoGarcia:2000}
\bibinfo{author}{\bibfnamefont{J.~J.} \bibnamefont{Brey}},
  \bibinfo{author}{\bibfnamefont{M.~J.} \bibnamefont{Ruiz-Montero}},
  \bibinfo{author}{\bibfnamefont{D.}~\bibnamefont{Cubero}}, \bibnamefont{and}
  \bibinfo{author}{\bibfnamefont{R.}~\bibnamefont{Garcia-Rojo}},
  \bibinfo{journal}{Physics of Fluids} \textbf{\bibinfo{volume}{12}},
  \bibinfo{pages}{876} (\bibinfo{year}{2000}).

\bibitem[{\citenamefont{P\"oschel and
  Brilliantov}(2003{\natexlab{b}})}]{PoeschelBrilliantov:2003}
\bibinfo{editor}{\bibfnamefont{T.}~\bibnamefont{P\"oschel}} \bibnamefont{and}
  \bibinfo{editor}{\bibfnamefont{N.~V.} \bibnamefont{Brilliantov}}, eds.,
  \emph{\bibinfo{title}{Granular Gas Dynamics}}, vol. \bibinfo{volume}{624} of
  \emph{\bibinfo{series}{Lecture Notes in Physics}}
  (\bibinfo{publisher}{Springer}, \bibinfo{address}{Berlin},
  \bibinfo{year}{2003}{\natexlab{b}}).

\end{thebibliography}

\end{document}